\documentclass[a4paper]{aa}
\usepackage{graphicx}
\usepackage{txfonts}
\usepackage{color}
\usepackage{natbib}
\usepackage{longtable}
\usepackage{hyperref}

\usepackage{amsmath}

\definecolor{blue}{RGB}{0,0,255}
\definecolor{red}{RGB}{255,0,0}
\definecolor{green}{RGB}{0,255,0}

\usepackage[normalem]{ulem}

\begin{document}

\title{One-dimensional, geometrically stratified semi-empirical models of the quiet-Sun photosphere and lower chromosphere} 

\author{J.M.~Borrero\inst{1} \and I.~Mili\'{c}\inst{1} \and A.~Pastor Yabar\inst{2}
\and A.~J.~Kaithakkal\inst{1} \and J.~de la Cruz Rodr{\'\i}guez\inst{2}}

\institute{Institut f\"ur Sonnenphysik, Sch\"oneckstr. 6, D-79104, Freiburg, Germany
\and
Institute for Solar Physics, Department of Astronomy, Stockholm University, AlbaNova University 
Centre, 10691 Stockholm, Sweden
}
\date{Recieved / Accepted}

\abstract{One-dimensional, semi-empirical models of the solar atmosphere are widely employed in numerous contexts within solar physics, ranging from the determination of element abundances and atomic parameters to studies of the solar irradiance and from Stokes inversions to coronal extrapolations. 
These models provide the physical parameters (i.e. temperature, gas pressure, etc.) in the solar atmosphere as a function of the continuum optical depth $\tau_{\rm c}$. 
The transformation to the geometrical $z$ scale (i.e. vertical coordinate) is provided via vertical hydrostatic equilibrium.}{Our aim is to provide updated, one-dimensional, semi-empirical models of the solar atmosphere
as a function of $z,$ but employing the more general case of three-dimensional magneto-hydrostatic equilibrium (MHS) instead of vertical hydrostatic equilibrium (HE).}
{We employed a recently developed Stokes inversion code that, along with non-local thermodynamic equilibrium effects, considers MHS instead of HE. This code
is applied to spatially and temporally resolved spectropolarimetric observations of the quiet Sun obtained with the CRISP instrument attached to the Swedish Solar Telescope.}
{We provide average models for granules, intergranules, dark magnetic elements, and overall quiet-Sun as a function of both $\tau_{\rm c}$ and $z$ from the photosphere to the lower chromosphere.}{We
demonstrate that, in these quiet-Sun models, the effect of considering MHS instead of HE is negligible. However, employing MHS increases the consistency of the inversion results
before averaging. We surmise that in regions with stronger magnetic fields (i.e. pores, sunspots, network) the benefits of employing the magneto-hydrostatic approximation will
be much more palpable.}

\titlerunning{Semi-empirical, geometrically stratified quiet-Sun models of the photosphere and chromosphere}
\authorrunning{Borrero et al.}
\maketitle

\def\kms{~km s$^{-1}$}
\def\deg{^{\circ}}
\def\df{{\rm d}}
\newcommand{\ve}[1]{{\rm\bf {#1}}}
\newcommand{\diff}{{\rm d}}

\section{Introduction}
\label{sec:introduction}

One-dimensional, semi-empirical models of the solar atmosphere are extremely important and useful for a variety of reasons. They have been used to determine solar abundances 
\citep{grevesse1989ti,blackwell1995fe,holweger1995fe,luis2002fe,borrero2008abun}
and transition probabilities in spectral lines \citep{gurtovenko1981gf,thevenin1989,thevenin1990,borrero2003atomic} by fitting 
the solar spectra. They can also be used to study the solar irradiance and its variation during the solar cycle
\citep{unruh1999,krivova2003}, or to assist in the extrapolation of magnetic fields from the photosphere toward the chromosphere and
corona \citep{thomas2013,thomas2015,nita2018,nita2023}. Finally, they are also very useful as initial guesses to initialise spectropolarimetric
inversions \citep{jaime2016,borrero2016pen,christoph2017,adur2018,christoph2019,adur2020,anabelen2021}, to investigate the formation of spectral lines 
using contribution and response functions \citep{bruls1992,borrero2017pen,ivan2017,vuda2022}, to study the turbulent solar magnetic field via
the Hanle effect \citep[e.g. ][]{ivan2012}, and to determine canonical values of radiative losses in the lower solar atmosphere
\citep{sobotka2016,abbasvand2020}.\\

The most widely used models correspond to average quiet-Sun models, such as HSRA \citep{gingerich1971} and HOLMU \citep{holweger1974}. Multi-component models, where the quiet Sun 
is split into average granular and average intergranular contributions, have also been presented \citep{borrero2002gra}. Available models of the magnetized Sun include average 
network and plage models \citep{solanki1986,solanki1992}, penumbra \citep{jc1994pen}, and umbra \citep{maltby1986,manolo1994}. A series of models for various solar features (quiet Sun, sunspots, plage, etc.)
{ have been presented by, among others, \citet{cristaldi2017}, as well as those commonly referred to} as VAL-models \citep{vernazza1981} and FAL-models \citep{fontenla1993,fontenla2009}.
All the aforementioned models provide the temperature, density, gas pressure, and sometimes the magnetic field and line-of-sight velocity, as a function of the continuum optical depth $\tau_{\rm c}$. While some of these
models only cover the photosphere ($\tau_{\rm c} \in [1,10^{-4}]$), others also include the chromosphere ($\tau_{\rm c} \in [10^{-4},10^{-8}]$) and even part of the transition region
($\tau_{\rm c} < 10^{-8}$). One common feature among all these one-dimensional semi-empirical models is that the $z$ scale is calculated using a gas pressure
obtained under the assumption of vertical hydrostatic equilibrium (HE). A better estimation of the gas pressure and, therefore, of the $z$ scale can be obtained if we apply
three-dimensional magneto-hydrostatic equilibrium (MHS) instead \citep{borrero2019mhs,borrero2021mhs}. Although this is particularly the case of regions where the magnetic field is strong (i.e. sunspot, pores, plages, etc.), as a first step,
we apply the aforementioned MHS method to the determination of one-dimensional semi-empirical models of the quiet Sun (granules, intergranules and dark magnetic element) here as a
function of both the continuum optical depth $\tau_{\rm c}$ and geometrical height $z$ for the photosphere and lower chromosphere: $\tau_{\rm c} \in [15,2.5 \times 10^{-6}]$. To this end, we perform MHS Stokes inversions 
of spectropolarimetric data at high spatial resolution recorded in the quiet Sun in two spectral lines: one photospheric (treated in local thermodynamic equlibrium) and one photospheric-chromospheric (treated 
under non-local thermodynamic equilibrium) in order to retrieve the physical properties of solar atmosphere in three dimensions ($T(x,y,z)$, $\vec{B}(x,y,z)$, $P_{\rm g}(x,y,z)$, etc.). Granules, intergranules, and 
dark magnetic elements will then be identified in the data, and average models as a function of $z$ and $\tau_{\rm c}$ will be calculated.\\

\begin{figure*}
\begin{tabular}{cc}
\includegraphics[width=8.45cm]{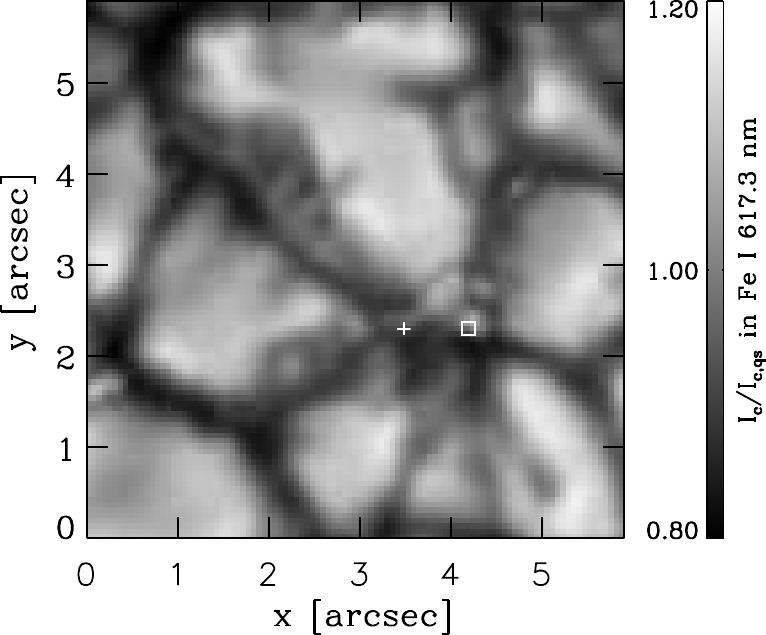} &
\includegraphics[width=8.45cm]{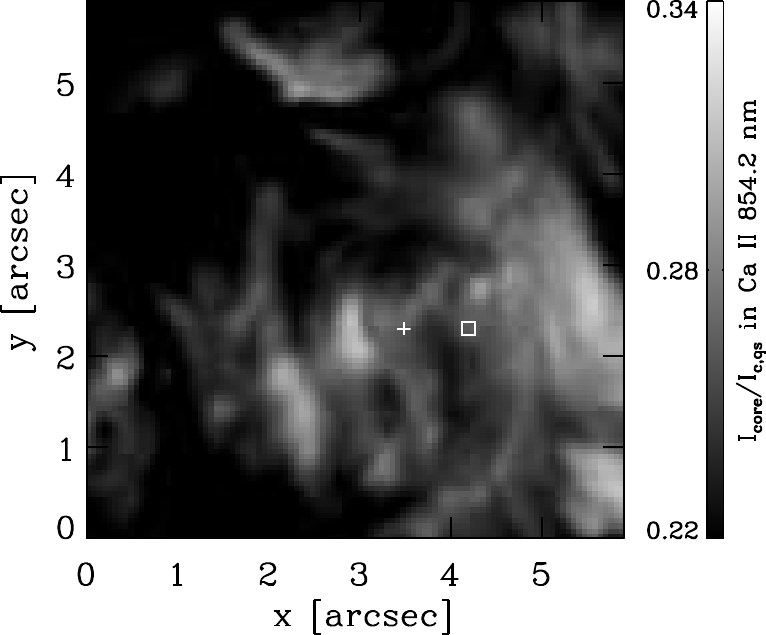} \\
\includegraphics[width=8.45cm]{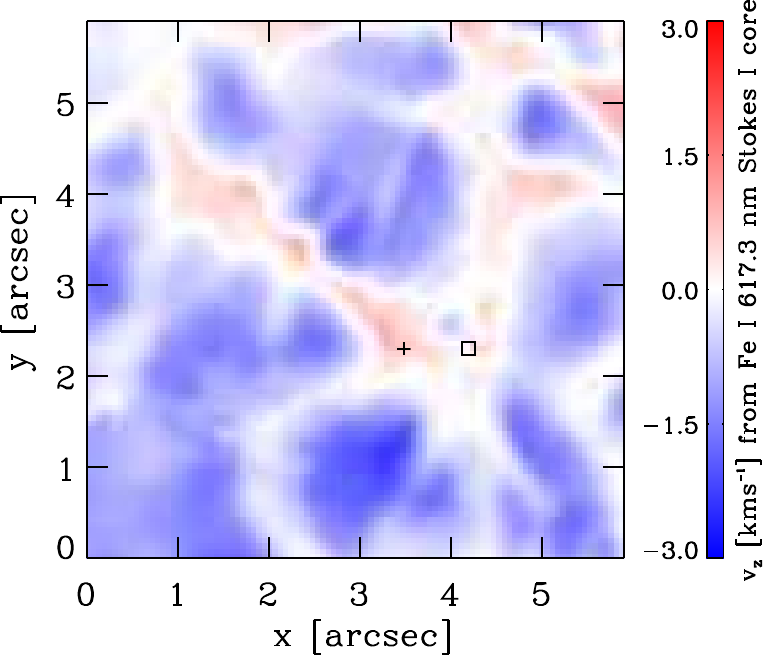} &
\includegraphics[width=8.45cm]{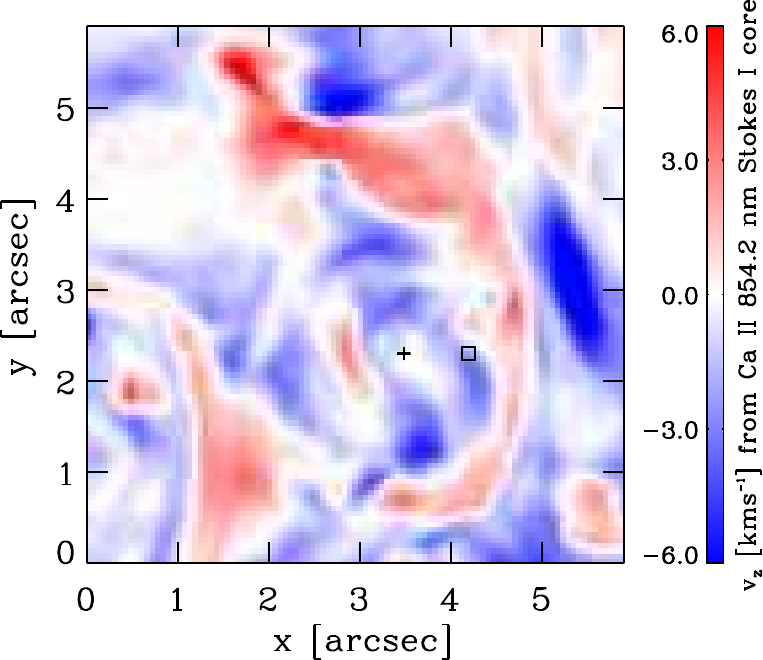} 
\end{tabular}
\caption{Example of the observations used in this work. {\it Top right}: Continuum intensity $I_{\rm c}$ in the \ion{Fe}{I} line at 617.3 nm
  normalized to the average value in the quiet Sun $I_{\rm c,qs}$ in the same line. {\it Top left}: Core intensity $I_{\rm core}$ in the \ion{Ca}{II}
  line at 854.2 nm, normalized to the average value in the quiet Sun $I_{\rm c,qs}$. {\it Bottom left}: Line-of-sight velocity obtained from the
  core of the \ion{Fe}{I} line at 617.3 nm. {\it Bottom right}: Line-of-sight velocity obtained from the core of the \ion{Ca}{II} line at 854.2 nm.
  Points indicated with the $+$ and $\square$ symbols are discussed in Section~\ref{sec:fits}.\label{fig:continuum}}
\end{figure*}

\section{Observations}
\label{sec:observations}

A region on the quiet-Sun was observed at the disk center ($\mu=1.0$) on April 24, 2019 with the CRisp Imaging SpectroPolarimeter 
\citep{scharmer2008crisp,alfred2021crisp} attached to the 1-m Swedish Solar Telescope \citep[SST;][]{goran2003sst}. Although the original data
are more exhaustive (i.e. longer time-sequence, larger field of view), here we only describe the portion
of the data that was used in our analysis. Other portions of the same dataset have previously been employed
in the study of a vortex tube \citep{fischer2020vortex} and of a quiet-Sun Ellerman bomb \citep{anjali2023}.\\

In our case, we employed the full observed Stokes vector, $\vec{I}_{\rm obs}=(I,Q,U,V)$, in the \ion{Fe}{I} line at 617.3 nm and the \ion{Ca}{II} line at 854.2 nm. The first line was recorded 
with a wavelength sampling of 35~m{\AA} across 15 spectral positions, from $-245$ to $+245$ m{\AA} with respect to the line center. The second 
line was recorded with a varying wavelength sampling, but it was later reinterpolated to a constant sampling of 100~m{\AA} 
in 17 spectral positions from $-800$ to $+800$ m{\AA} with respect to the line center. In addition to this, for the \ion{Ca}{II} line, an 
additional wavelength point at $+2.4$~{\AA} from the line center is used to normalise the spectra. This spectral line features such 
extended wings that this wavelength is not far enough away to guarantee that the true continuum is reached. In fact, at this wavelength, the Fourier Transform Spectrometer 
\citep[FTS; ][]{fts1987} atlas intensity reaches only 0.76, which is the number employed for normalization purposes. Table~\ref{tab:atomic} presents the atomic parameters 
pertaining to these two spectral lines.\\

The field of view in our dataset includes 100 pixels along each spatial dimension on the solar surface with a spatial sampling
of 0.059 arcsec ($\approx 44$ km). The total field-of-view of $5.9 \times 5.9$ arcsec$^2$ is sufficient to cover several granular cells' 
and intergranules' lanes. The data calibration was carried out using the SSTRED pipeline \citep{jaime2015crisp,lofdahl2021crisp} and was further processed 
with the multi-object, multi-frame blind deconvolution code \citep[MOMFBD,][]{lofhadl2002,michiel2005momfbd}. In our work, we used a time series of
28 snapshots with a sampling of roughly 30~seconds, which thus spanned about 15 minutes of solar evolution.\\

The spatial resolution of the data is about 0.12 arcsec, and the noise level in the polarization signals is roughly $2 \times 10^{-3}$
in units of the continuum intensity. For context we provide, in Figure~\ref{fig:continuum}, images of the continuum intensity in the
\ion{Fe}{I} line (top left) and core intensity in the \ion{Ca}{II} line (top right) for one of the 28 snapshots in our dataset. 
In addition, the line-of-sight velocities (or $v_z$) obtained from the line core of both spectral lines are also provided in this figure: 
\ion{Fe}{I} (bottom left), \ion{Ca}{II} (bottom right). The line-of-sight velocities were obtained by performing a second-degree polynomial 
fit in three points centered around the wavelength point with the lowest intensity.\\

\begin{table*}
\caption{Atomic parameters for spectral lines employed in our work.}
\label{tab:atomic}
\centering                 
\begin{tabular}{ccccccccccc}
\hline
\hline
Atom & Ionization & $\lambda_0$ & $n_\lambda$ & $\Delta\lambda$ & ${\rm e}_{\rm low}$ & ${\rm e}_{\rm upp}$ & 
$\log_{10}{{\rm g}\,{\rm f}}$ & ${\rm E}_{\rm low}$ & $\alpha$ & $\sigma/{\rm a}_{0}^{2}$ \\
 & & $\textrm{[nm]}$ & & $\textrm{[m{\AA}]}$ & & & & \textrm{[eV]} & & \\
\hline
Fe & {\sc I} & 617.334 & 15 & 35 & $^5$P$_1$ & $^5$D$_0$ & -2.880 & 2.223 & 0.266 & 280.6\\
Ca & {\sc II} & 854.209 & 17 & 100 & $^2$D$_{5/2}$ & $^2$P$_{3/2}$ & -0.360 & 1.700 & 0.275 & 291.0\\
\hline                                           
\end{tabular}
\tablefoot{$\lambda_{0}$ is the central wavelength for the electronic transition associated with the spectral line; $n_{\lambda}$ 
is the number of observed wavelengths for each spectral line; $\Delta\lambda$ is the spectral sampling in m{\AA}; 
${\rm e}_{\rm low}$ and ${\rm e}_{\rm upp}$ are the electronic configurations of the lower and upper energy level, respectively; 
${\rm E}_{\rm low}$ is the excitation potential (in eV) of the lower energy level. These atomic data are adopted from NIST \citep{NIST_ASD}. In particular,
\citet{nave1994} is used for Fe and \citet{edlen1956ca} for Ca. The $\alpha$ and $\sigma/{\rm a}_{0}^{2}$ are the velocity exponent 
and collision cross-section parameters (in units of Bohr's radius, $a_0$), respectively, as defined in the Anstee, Barklem, 
and O'Mara collision theory for the broadening of metallic lines by neutral hydrogen collisions \citep{anstee1995, barklem1997, barklem1998}.}
\end{table*}

\section{Stokes inversion}
\label{sec:inversion}

The observed Stokes vector $\vec{I}_{\rm obs}=(I,Q,U,V)$ as a function of wavelength in the two aforementioned spectral lines is inverted at each spatial position $(x,y)$ 
on the solar surface with the FIRTEZ Stokes inversion code \citep{adur2019invz}. { This code is currently the only Stokes inversion code capable of employing MHS equilibrium
to determine the gas pressure and geometrical height $z$ in the solar atmosphere}. For this dataset, we performed six inversion cycles. The reason we needed
multiple cycles is explained in Sect.~\ref{sec:convergence}. The configuration of each inversion cycle is summarized in Table~\ref{tab:inversion}. During each cycle, the 
$\chi^{2}$ merit function between the synthetic $\vec{I}_{\rm syn}$ and observed Stokes vector $\vec{I}_{\rm obs}$ is minimized using, as a starting atmospheric model,
the results from the previous cycle. At each spatial position $(x,y),$ the atmospheric model is discretized along 128 points in the vertical $z$ direction with a step size 
of $\Delta z = 12$~km.\\

During the inversion, the \ion{Fe}{I} line is always treated under local thermodynamic equilibrium (LTE), while the \ion{Ca}{II} line is treated under non-local thermodynamic equilibrium (NLTE) via departure coefficients \citep{anjali2023} in the line opacity and source function \citep{hector1998,hector2000}. The departure coefficients for the upper, $\beta_{\rm upp}$, 
and lower, $\beta_{\rm low}$, atomic levels are defined as the ratio between the level populations in NLTE and LTE. Although $\beta_{\rm upp}$ and $\beta_{\rm low}$ for the \ion{Ca}{II} 
line are kept unaltered during all iterations of a given cycle \citep[i.e. akin to the so-called fixed departure coefficients approximation;][]{basilio2022}, they are updated at the end of each cycle. 
This is done at every point of the full $(x,y,z)$ domain using the SNAPI code \citep{ivan2017,ivan2018} and assuming complete redistribution in frequencies. SNAPI solves the statistical 
equilibirum equations with a \ion{Ca}{II} atomic model that includes five bound states and one continuum level \citep{shine1974ca}. Collisional rates are determined using NLTE electron densities
and employing a hydrogen model with five bound states plus one continuum level \citep{ivan2021}. Since FIRTEZ only uses LTE to determine the electron density \cite{saha1920}, we modified
it to consider a new departure coefficient, $\beta_{\rm e}$, which is defined as the ratio between the electron density in NLTE and in LTE, so that FIRTEZ can also employ NLTE electron densities
when calculating the continuum opacity, mean molecular weight, and gas pressure (i.e. equation of state) during the inversion. Unlike $\beta_{\rm upp}$ and $\beta_{\rm low}$, $\beta_{\rm e}$ is not updated after each 
inversion cycle, but instead kept constant to those of the initial guess model. The reasons for this, along with its implications, are discussed in Sect.~\ref{sec:beta_elec}.\\

During cycles one through three, vertical hydrostatic equilibrium is assumed to determine the gas pressure $P_{\rm g}$. At this point, only Stokes $I$ is inverted. This is done by giving 
zero weight in $\chi^2$ to the other Stokes parameters: $w_i=1$, $w_q=w_u=w_v=0$. The number of free parameters (i.e. nodes) are eight for the temperature and eight for the vertical velocity. 
In order to avoid large grid-to-grid variations in the $z$ direction, we used a Tikhonov regularization \citep{tikhonov1995} whereby solutions with large $z$-derivatives of $T(z)$ and $v_z(z)$
are penalized in the $\chi^2$ merit-function \citep{jaime2019}.\\

To initialise the first cycle, an atmospheric model estimation needs to be constructed. In our case, we used the temperature $T(z)$ and gas pressure $P_{\rm g}(z)$ 
from the VALC model \citep{vernazza1981}. The initial departure coefficients $\beta_{\rm upp}(z)$, $\beta_{\rm low}(z),$ and $\beta_{\rm e}(z)$ are also calculated for this model.
The same values are used at each $(x,y)$ position. Vertical velocities $v_z$ are initialized differently for each $(x,y)$ and determined via the 
center-of-gravity method applied to the intensity profile (i.e. Stokes $I$). The value of $v_z$ thus obtained is assumed to be constant with $z$.\\

In cycles four through six we fit, along Stokes $I$, the remaining Stokes parameters ($Q$, $U,$ and $V$). Here, the weights in the $\chi^2$ merit function given to the linearly polarized Stokes parameters are slowly increased
from cycle to cycle (see Table~\ref{tab:inversion}). In order to fit all four Stokes parameters, we include nodes in $B_x$, $B_y$, and $B_z$. During cycle four, we allow for
one node in each of the components of the magnetic field. This is increased to two nodes for each component during cycles five and six (see Table~\ref{tab:inversion}). The values for $B_{x}$, $B_{y,}$ and $B_{z}$ 
are initialized at the start of the fourth inversion cycle, at each spatial location $(x,y)$, via the weak-field approximation \citep{jefferies1991} and, to begin with, assumed to be constant in $z$. 
The horizontal components of the magnetic field $B_x$ and $B_y$ are regularized employing the Tikhonov regularization, but using a regularization function in $\chi^2$ that penalises 
deviations from the zero value instead of the vertical derivative (as was the case of $T(z)$ and $v_z(z)$ during cycles one through three). With this we try to avoid obtaining large values of the 
horizontal component of the magnetic field produced by the inversion of low signal-to-noise $Q$ and $U$ profiles that are often seen in quiet-Sun spectropolarimetric observations 
\citep{borrero2011qs,borrero2012qs}. It is important to note that because the magnetic field is being inverted during cycles four through six and because the gas pressure is also being changed, the 
fits to Stokes $I$ obtained during the first three cycles can change. In order to make sure that good fits to Stokes $I$ are still obtained during these cycles, we also need to include $T$ and $v_z$ during 
the inversion. This is done, again, with eight nodes in each of these two physical parameters.\\

Owing to the existence of a non-zero magnetic field in cycles four through six, it is now possible to obtain the gas pressure $P_{\rm g}$ while accounting for the Lorentz-force term 
\citep[magneto-hydrostatic equilibrium; ][]{borrero2019mhs} instead of assuming vertical hydrostatic equilibrium as in cycles one through three. To this end, we employed the procedure described in 
detail in \citet[][]{borrero2021mhs} (see their Fig.~2) and employed the following boundary conditions: $P_{\rm g}(z=0) = 2.755 \times 10^5$ dyn~cm$^{-2}$ and $P_{\rm g}(z=z_{\rm max}) = 1.675$ dyn~cm$^{-2}$.
These boundary conditions apply for every $(x,y)$ pixel. The gas pressure $P_{\rm g}$ thus obtained is more realistic than the hydrostatic values, and it allows for a reliable inference of the
physical parameters (temperature, vertical velocity, and so on) as a function of the Cartesian coordinate $z$ (i.e. geometrical height) instead of using the optical depth $\tau_{\rm c}$.\\

\begin{table}
\caption{Inversion setup. In type 1 regularization, the $\chi^2$ merit-function is penalized with the vertical gradient of the physical parameter that is being regularized: i.e. $\df T(z)/ \df z$.
In type 2 regularization, the $\chi^2$ merit function is penalized with deviations of physical parameters from the zero value.}
\label{tab:inversion}
\begin{center}
\begin{tabular}{c|cccc}
\hline
cycles & 1-3 & 4 & 5 & 6 \\
\hline
$P_{\rm g}$ & HE & MHS & MHS & MHS\\
$w_i$ & 1 & 1 & 1 & 1 \\
$w_{q,u}$ & 0 & 1 & 1.5 & 2.5\\
$w_{v}$ & 0 & 1 & 1 & 2\\
nodes $T(z)$ & 8 & 8 & 8 & 8\\
regul $T(z)$ & type 1 & type 1 & type 1 & type 1 \\
nodes $v_z(z)$ & 8 & 8 & 8 & 8\\
regul $v_z(z)$ & type 1 & type 1 & type 1 & type 1\\
nodes $B_z(z)$ & 0 & 1 & 2 & 2\\
regul $B_z(z)$ & no & no & no & no\\
nodes $B_x(z)$/$B_y(z)$ & 0 & 1/1& 2/2 & 2/2\\
regul $B_x(z)$/$B_y(z)$ & no & type 2 & type 2 & type 2\\
\hline
\end{tabular}
\end{center}
\end{table}

\section{Inversion results}

\subsection{Convergence of Non-LTE atomic departure coefficients}
\label{sec:convergence}

\begin{figure}
\includegraphics[width=8cm]{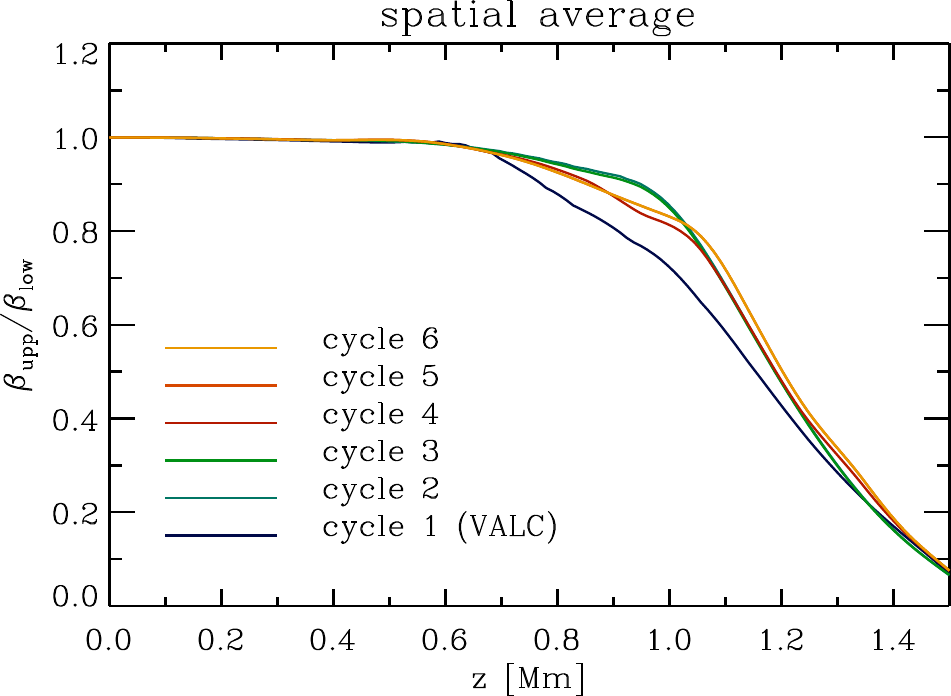}
\caption{Vertical $z$ stratification of spatial $(x,y)$ average of ratio between the upper level, $\beta_{\rm upp}$, and lower level, $\beta_{\rm low,}$ departure coefficients. Color-codes indicate the values
after each inversion cycle. The initial cycle (dark blue) corresponds to the departure coefficients obtained from the VALC model.\label{fig:beta}}
\end{figure}

In the previous section, we mention that out of the six inversion cycles performed on the data, the first three were carried out with gas pressure $P_{\rm g}$ resulting from
vertical hydrostatic equilibrium, whereas in cycles four through six, the gas pressure $P_{\rm g}$ was obtained via three-dimensional magneto-hydrostatic equilibrium (see also Table~\ref{tab:inversion}).
The reason as to why three cycles are needed under each approximation is because at the end of each inversion cycle the departure coefficients are updated, which, in turn, changes the best-fit Stokes profiles
and thus also the inferred physical parameters. It takes three cycles of updating the departure coefficients for this procedure to converge.
In order to demonstrate this convergence, in Fig.~\ref{fig:beta} we present the spatial $(x,y)$ average of the ratio between the upper-level departure
coefficient, $\beta_{\rm upp}$, and the lower-level departure coefficient $\beta_{\rm low}$ as a function of the vertical Cartesian coordinate $z$ (i.e. geometrical height) at each inversion cycle.
At cycle 1, the departure coefficients correspond to those of the VALC model, and they are the same at every spatial $(x,y)$ pixel because the inversion was initialized with this model everywhere.\\

The reason for illustrating this ratio is because, at visible and near-infrared wavelengths, the source function $S(T[z])$, where $T[z]$ refers to the temperature dependence along $z$, is proportional 
to the ratio $\beta_{\rm upp}(z) / \beta_{\rm low}(z)$. If this ratio remains unchanged in two consecutive inversion cycles it means the inversion can be stopped because no further updates are needed to the 
departure coefficients and, hence, the temperature $T(z)$. In the case of hydrostatic equilibrium, we reach this convergence after only three cycles, as bespoken by the fact the $\beta_{\rm upp}/\beta_{\rm low}$ is almost 
identical after cycles two and three.\\

Once convergence has been achieved, vertical hydrostatic equilibrium is switched off and three-dimensional magneto-hydrostatic equilibrium is used instead. Right at the end of inversion cycle 4, we already have
an estimation of the magnetic field $\vec{B}$, and, consequently, the gas pressure is calculated under MHS. This of course differs from the gas pressure under HE, thereby yielding very different level 
populations and thus leading to a new sudden change in the departure coefficients (compare the green curve (cycle three) with the red curve (cycle four) in Fig.~\ref{fig:beta}). Therefore, a new series of inversion 
cycles under MHS need to be performed until convergence is again achieved. This occurs in cycle six, as departure coefficients after cycle five (orange) and six (yellow) are very similar.\\

\subsection{FIRTEZ equation of state}
\label{sec:beta_elec}

\begin{figure*}
\begin{tabular}{cc}
\includegraphics[width=8cm,bb= 70 370 550 715]{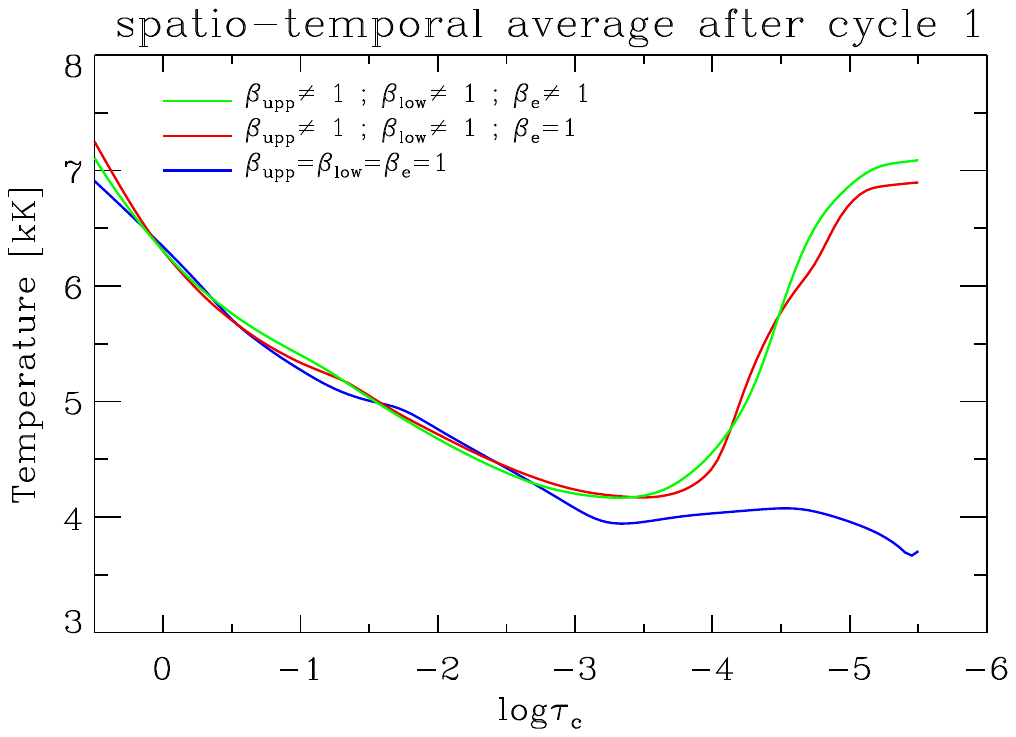} &
\includegraphics[width=8cm,bb= 70 370 550 715]{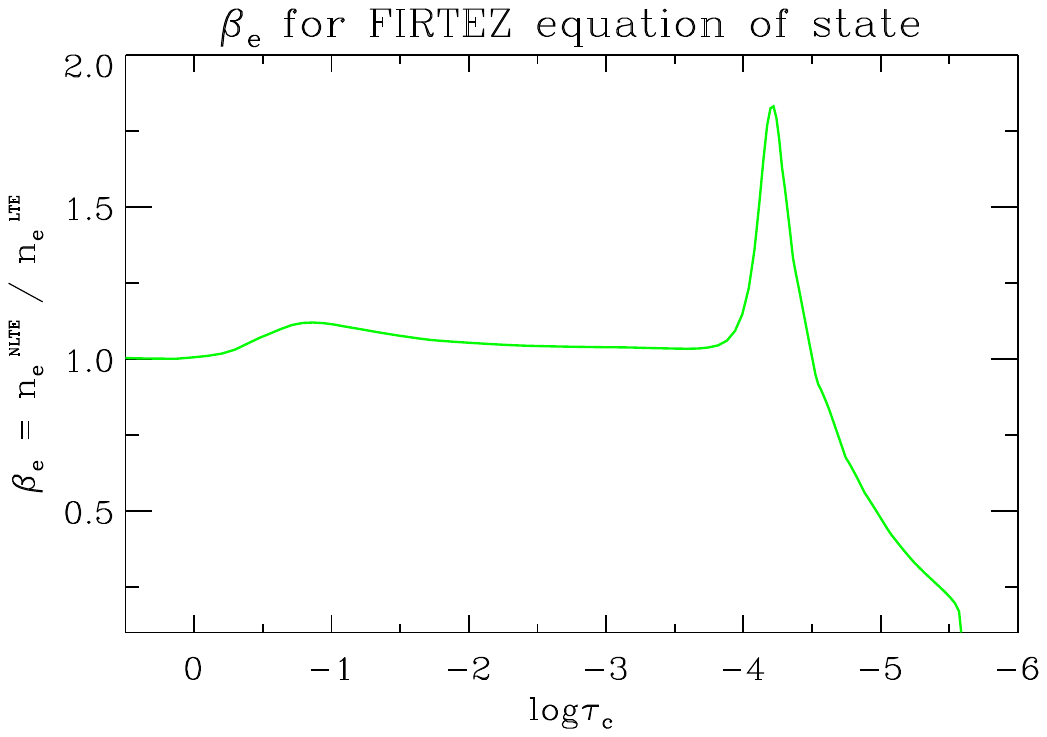} 
\end{tabular}
\caption{Effects of the equation of state in the retrieval of the temperature. {\it Left panel}: Spatially and temporally averaged
  $T(\log\tau_{\rm c})$ obtained after inversion cycle 1 employing (blue) inversion under LTE; (red) inversion employing NLTE atomic level departure
  coefficients and LTE equation of state in FIRTEZ; (green) inversion employing NLTE atomic level departure coefficients and NLTE equation of state
  in FIRTEZ. {\it Right panel}: Departure coefficient for electron density $\beta_{\rm e}(\log\tau_{\rm c})$ used in FIRTEZ equation of state in the
  third case presented in the left panel.\label{fig:beta_elec}}
\end{figure*}

As already explained in Sect.~\ref{sec:inversion}, the atomic level departure coefficients $\beta_{\rm upp}$ and $\beta_{\rm low}$ are determined, using SNAPI, with a NLTE electron density. This is important
because $\beta_{\rm upp}$ and $\beta_{\rm low}$ calculated using NLTE electron densities can differ significantly from those calculated using LTE electron densities (see Pastor Yabar et al. {\it in preparation}). 
However, this introduces an inconsistency with FIRTEZ since this code calculates the electron density using Saha's equation \cite[i.e. LTE][]{saha1920}. In order to remove this inconsistency, FIRTEZ has been 
modified to account for deviations from local thermodynamic equilibrium in the calculation of the electron density via a corresponding departure coefficient, referred to as $\beta_{\rm e}$. Unlike the 
case of the atomic level departure coefficients $\beta_{\rm upp}$ and $\beta_{\rm low}$, the departure coefficients for the electron density $\beta_{\rm e}$ are calculated only once for the VALC model and 
kept the same for all $(x,y)$ pixels in the field of view and for all inversion cycles described in Sect.~\ref{sec:inversion} 
(see also Table~\ref{tab:inversion}). The reason for this is that the inclusion of $\beta_{\rm e}$ in the determination of the temperature via the Stokes inversion has a minor effect compared to the
inclusion of $\beta_{\rm upp}$ and $\beta_{\rm low,}$ and therefore it is much more important to update the latter than the former. This procedure does not fully remove the aforementioned inconsistency, but,
judging from the effect that $\beta_{\rm e}$ has on the temperature inference, we gauge that our systematic error is, at most, of the order of 300~K, which is probably below the systematic uncertainties of
chromospheric inversions. In order to illustrate this, in Fig.~\ref{fig:beta_elec} (left panel) 
we present the average temperature stratification obtained from the inversion, for all pixels and all times at the end of cycle 1 in three different cases: { (a)} pure LTE inversion (blue; 
$\beta_{\rm e} = \beta_{\rm upp} = \beta_{\rm low} =1$); { (b)} atomic level departure coefficients $\beta_{\rm upp}$ and $\beta_{\rm low}$, obtained with NLTE electron density, but Stokes inversion using LTE electron 
density (red; $\beta_{\rm e} = 1$ ; $\beta_{\rm upp} \ne 1$ ; $\beta_{\rm low}  \ne 1$); and, finally, { (c)} atomic level departure coefficients $\beta_{\rm upp}$ and $\beta_{\rm low}$, obtained with NLTE electron density and 
Stokes inversion also using NLTE electron density (green; $\beta_{\rm e} \ne 1$ ; $\beta_{\rm upp} \ne 1$ ; $\beta_{\rm low}  \ne 1$). As can be seen, the largest 
correction to the temperature ($ \approx 1000-3000$~K) comes from considering NLTE level populations. Including NLTE electron density in the FIRTEZ equation of state typically yields slightly larger temperatures 
than using LTE electron density, but the differences are contained within 200-300~K. For completeness, Fig.~\ref{fig:beta_elec} (right panel) provides 
$\beta_{\rm e}$ as a function of $\tau_{\rm c}$ as determined for the VALC model and used in this work.\\

\subsection{Maps of the physical parameters}
\label{sec:maps}

\begin{figure*}
\begin{tabular}{cc}
\includegraphics[width=8cm]{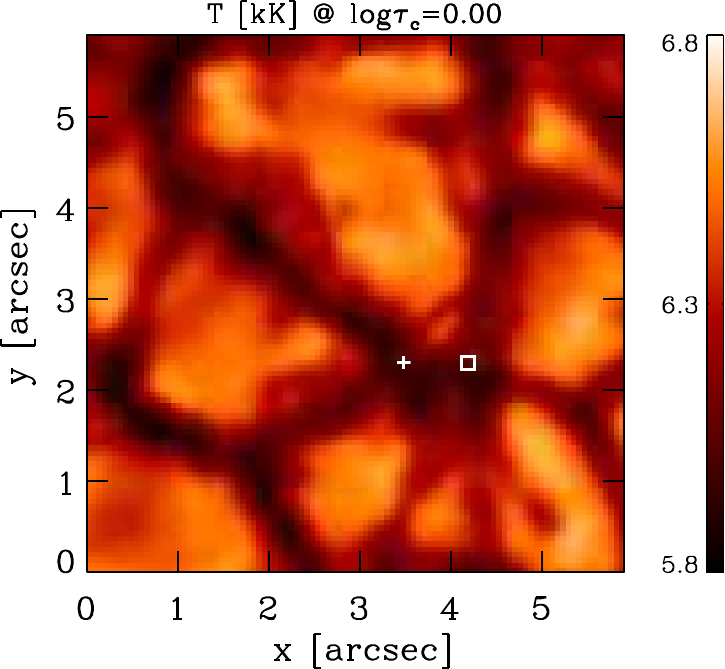} &
\includegraphics[width=8cm]{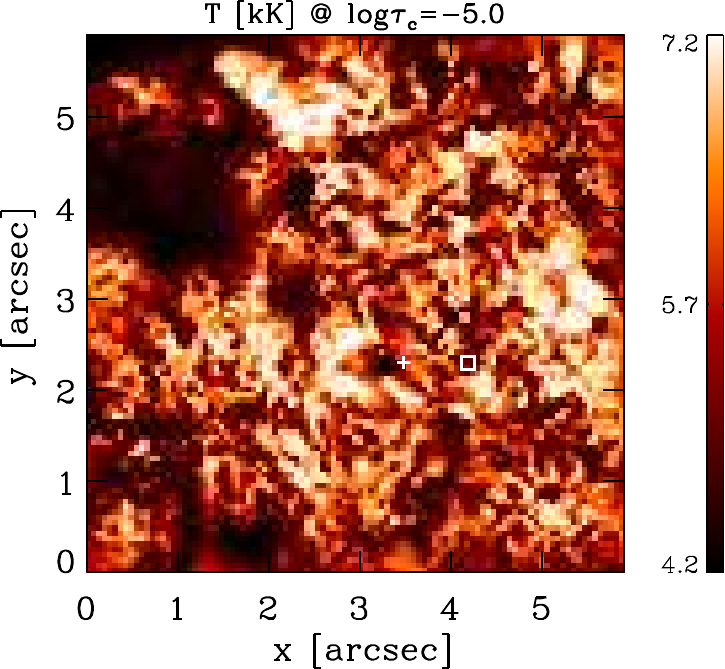} \\
\includegraphics[width=8cm]{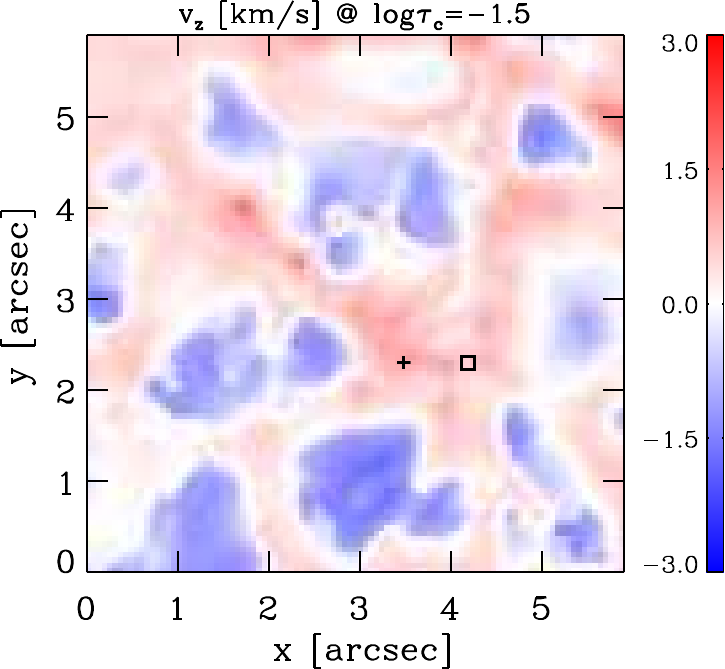} &
\includegraphics[width=8cm]{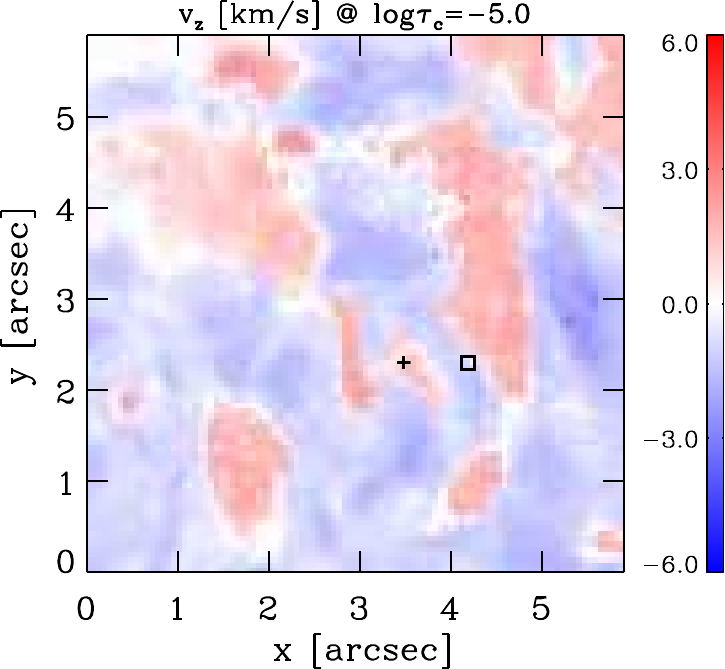}
\end{tabular}
\caption{Physical parameters resulting from inversion of Stokes profiles of one of the snapshots described in Sect.~\ref{sec:observations}. {\it Top panels}: Temperature $T(x,y)$ at $\log\tau_{\rm c}=0$ (left) and $\log\tau_{\rm c}=-5$ (right). {\it Bottom panels}: Vertical component of velocity $v_z(x,y)$ at $\log\tau_{\rm c}=-1.5$ (left) and $\log\tau_{\rm c}=-5$(right).\label{fig:phys1}}
\end{figure*}

In order to illustrate the reliability of the inversion process described in Sect.~\ref{sec:inversion}, in this section we present several maps of the physical parameters on the solar surface $(x,y)$
at different optical depths\footnote{Although our models are originally inferred in the three-dimensional $(x,y,z)$ Cartesian domain, it is better to perform the comparison with actual observables in
the $\tau_{\rm c}$ scale.}. The top panels of Figure~\ref{fig:phys1} show maps of the temperature at $\log\tau_{\rm c}=0$ (i.e. low photopshere) and $\log\tau_{\rm c}=-5$ (i.e. low chromosphere). These two optical
depths correspond, approximately, to the height of formation of the continuum in the \ion{Fe}{I} line and the core in the \ion{Ca}{II} line, respectively, and therefore can
be readily compared with the top panels in Fig.~\ref{fig:continuum}. The correlation between continuum intensity in \ion{Fe}{I} and temperature at $\log\tau_{\rm c}=0$ is very clear. Correlations between core 
intensity in \ion{Ca}{II} and temperature at $\log\tau_{\rm c}=-5$ is lower, albeit still present. This happens because the source function couples more strongly to the local temperature in the photosphere than 
in the chromosphere, as demonstrated by the fact that the ratio between the upper and lower departure coefficients is very close to unity in the photosphere (see Fig.~\ref{fig:beta}).\\

\begin{figure*}
\begin{tabular}{cc}
\includegraphics[width=8cm]{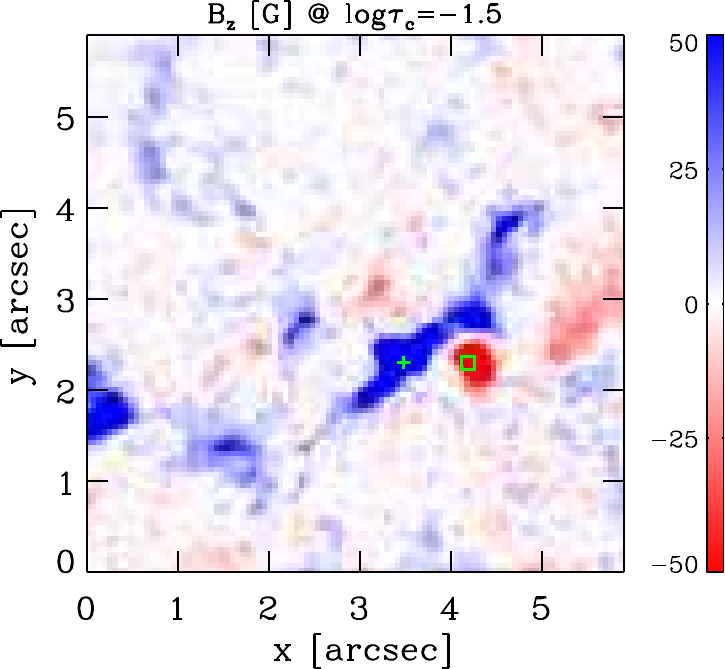} &
\includegraphics[width=8cm]{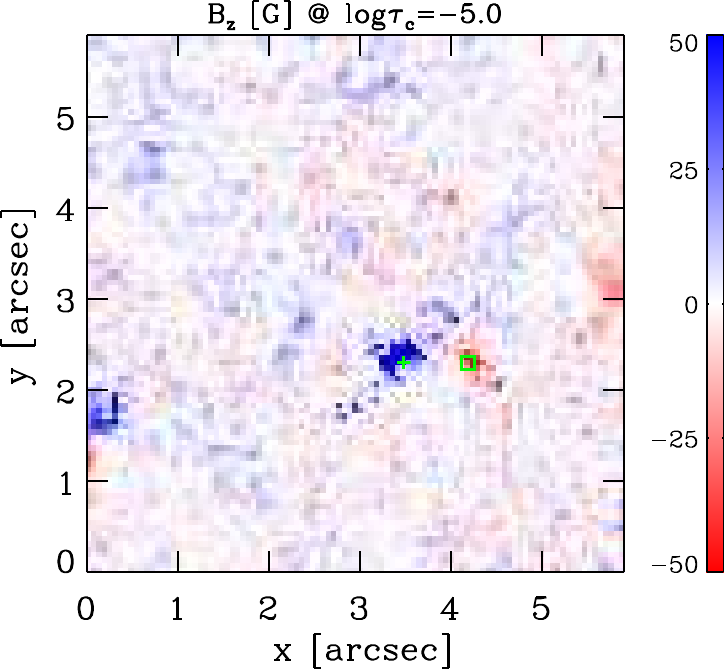} \\
\includegraphics[width=8cm]{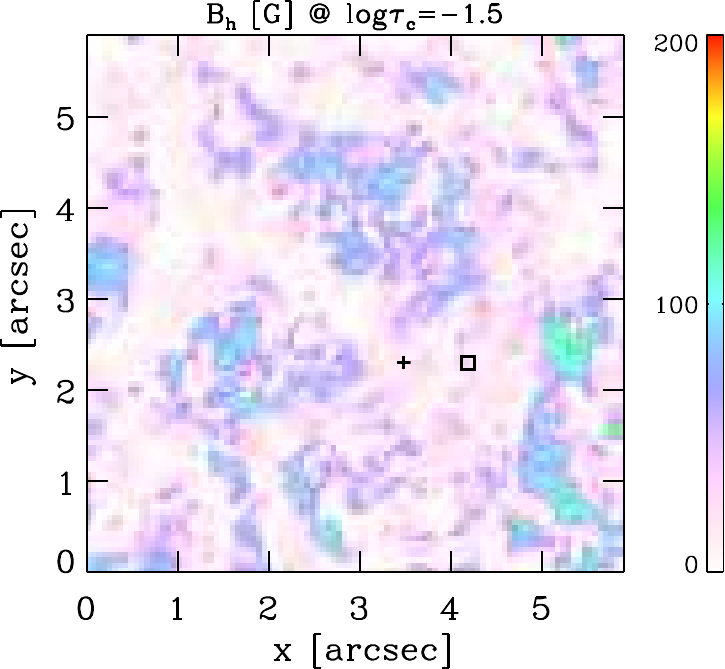} &
\includegraphics[width=8cm]{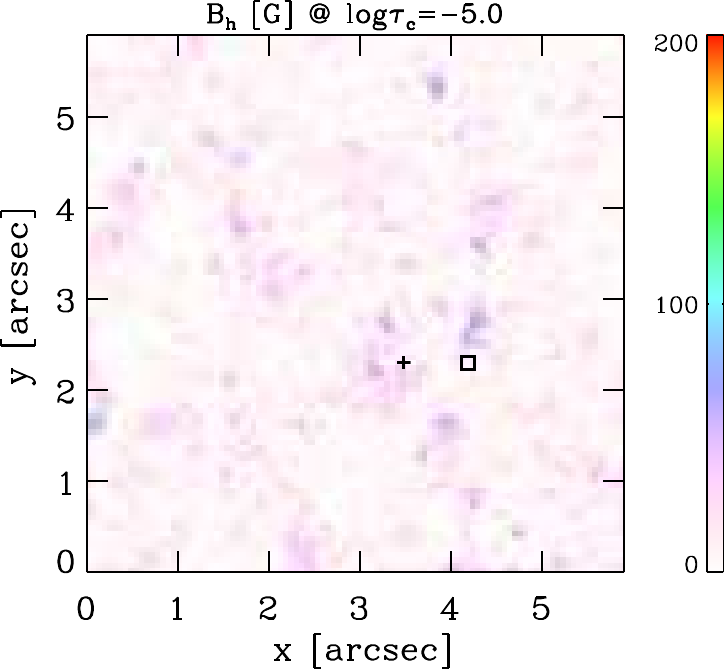}
\end{tabular}
\caption{Physical parameters resulting from inversion of Stokes profiles of the same snapshot as in Fig.~\ref{fig:phys1}. {\it Top panels}: Vertical component of magnetic field $B_z(x,y)$ at $\log\tau_{\rm c}=-1.5$ (left) and $\log\tau_{\rm c}=-5$ (right). {\it Bottom panels}: Horizontal component of magnetic field $B_h(x,y)$ at $\log\tau_{\rm c}=-1.5$ (left) and $\log\tau_{\rm c}=-5$ (right).\label{fig:phys2}}
\end{figure*}

The bottom panels in Fig.~\ref{fig:phys1} show the vertical (line-of-sight) velocities $v_z$ at optical depths of $\log\tau_{\rm c}=-1.5$ (i.e. mid photosphere) and $\log\tau_{\rm c}=-5$ (i.e. low chromosphere).
Again, these two optical depths correspond to the height of formation of the core in the \ion{Fe}{I} line and the core in the \ion{Ca}{II} line, respectively. Consequently, they can be compared with the
bottom panels in Fig.~\ref{fig:continuum}. A good correlation is also present. Inversion results for the vertical and horizontal components of the magnetic 
field, $B_z$ and $B_h=\sqrt{B_x^2+B_y^2}$, at two optical depths ($\log\tau_{\rm c}=-1.5,-5$) are presented in Fig.~\ref{fig:phys2}. As expected from the fact that the polarization signals are much lower in the 
\ion{Ca}{II} line than in the \ion{Fe}{I} line, these maps confirm that the magnetic field strength decreases with height. We also note that the horizontal component of the magnetic field at $\log\tau_{\rm c}=-1.5$
is concentrated mostly inside granular cells.\\

\subsection{Sample fits to the observations}
\label{sec:fits}

\begin{figure*}
\begin{tabular}{cc}
\includegraphics[width=8cm,bb=-14 0 341 200]{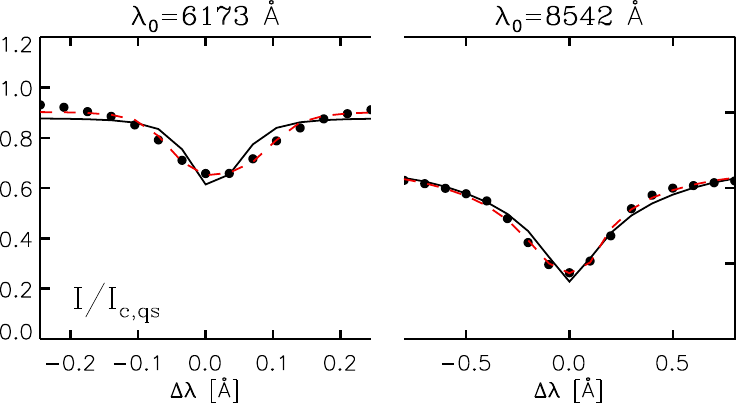} &
\includegraphics[width=8cm,bb=-14 0 341 200]{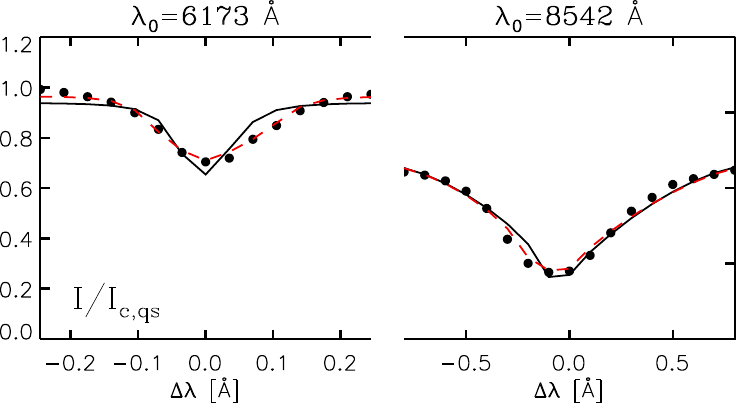} \\
\includegraphics[width=8cm,bb=0 0 355 200]{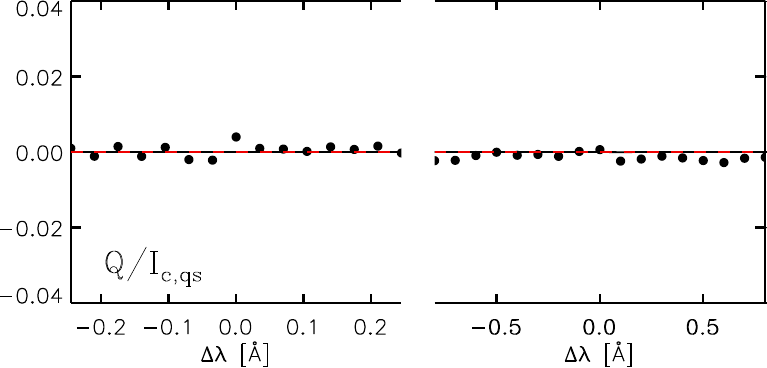} &
\includegraphics[width=8cm,bb=0 0 355 200]{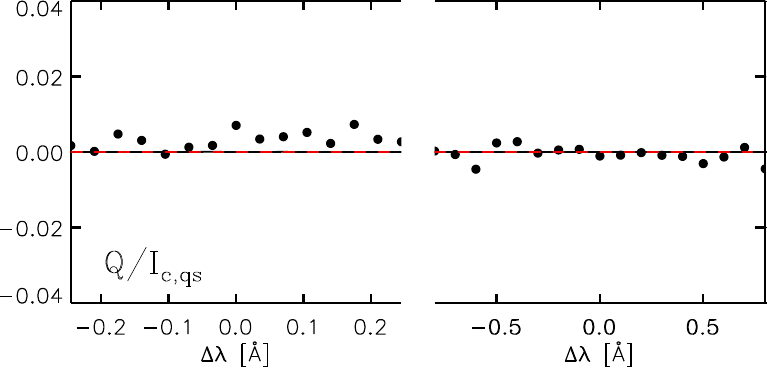}\\
\includegraphics[width=8cm,bb=0 0 355 200]{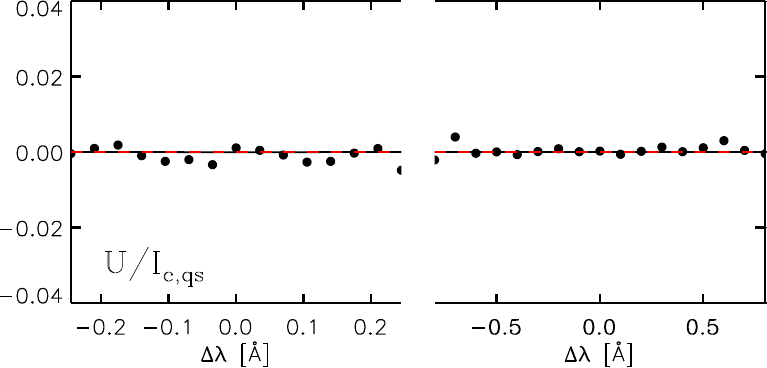} &
\includegraphics[width=8cm,bb=0 0 355 200]{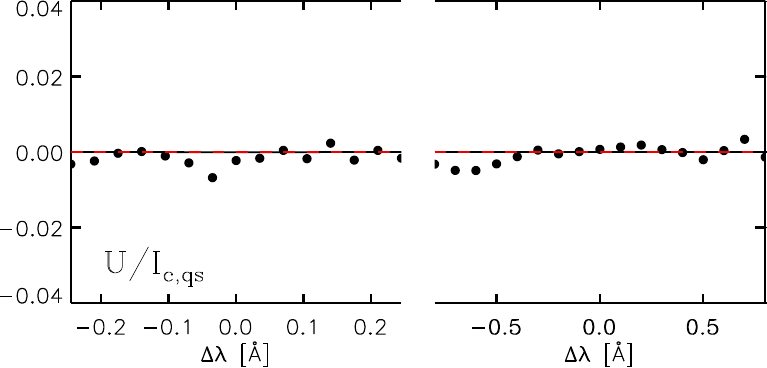} \\
\includegraphics[width=8cm,bb=0 0 355 200]{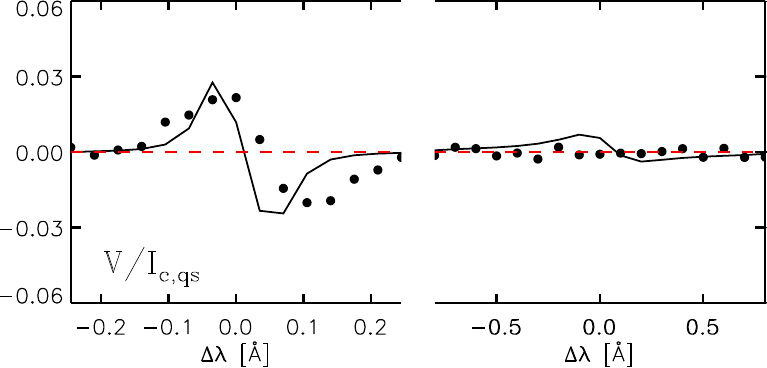} &
\includegraphics[width=8cm,bb=0 0 355 200]{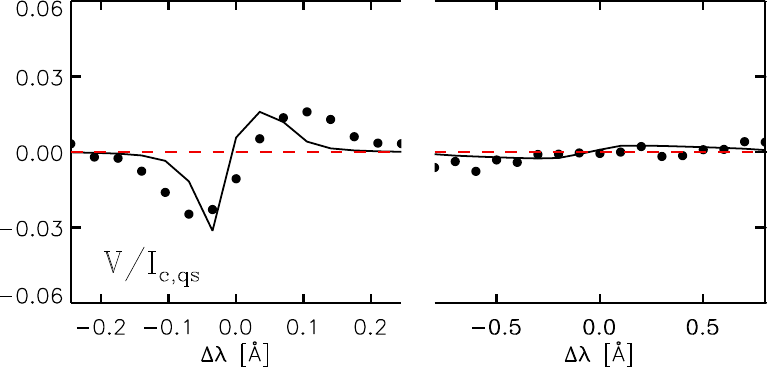}
\end{tabular}
\caption{Observed (circles) and fitted Stokes profiles in two examples indicated in Fig.~\ref{fig:continuum}. {\it Left panels}: Results corresponding to pixel indicated by plus ($+$) symbol. {\it Right panels}: Results corresponding to pixel indicated by the square ($\square$). {\it From top to bottom}: Intensity (Stokes $I$), linear polarization (Stokes $Q$ and $U$), and circular polarization (Stokes $V$). Results after cycle 3 are indicated by the dashed red lines, whereas results after cycle 6 are shown as solid black lines.\label{fig:fits}}
\end{figure*}

Two examples of the observed (circles), fitted after the cycle 3 (red dashed lines) and fitted after cycle 6 (solid black lines) Stokes profiles, are given in Figure~\ref{fig:fits}. These correspond to 
the plus ($+$) and square ($\square$) symbols in Figs.~\ref{fig:continuum},~\ref{fig:phys1}, and ~\ref{fig:phys2}, and are located in regions of large positive ($B_z >0$) and negative ($B_z <0$) polarities.
The fits to Stokes $I$ are excellent after cycle 3, but they degrade somewhat after cycle 6, because the polarization profiles are also inverted and are given larger weights than Stokes $I$ in order to
infer the three components of the magnetic field. One possibility that was explored, in order to improve to fits to Stokes $I$, was to fix the temperature $T(x,y,z)$ to those obtained in cycle 3 for the cycles that followed.
Unfortunately, this produced even worse fits than the ones presented here in solid black lines because the gas pressure $P_{\rm g}(x,y,z)$ is being changed despite the temperature being kept the same, resulting in very 
different $T(x,y,\log\tau_{\rm c})$. Fortunately, the spatially average temperature stratification as a function of optical depth $T(\tau_{\rm c})$ is very similar after cycles 3 and 6, meaning that the errors in the 
inference of the temperature in the latest cycle that might arise from the misfits to Stokes $I$ cancels out after averaging.\\

Regarding the polarization signals, we deem the fits to Stokes $Q$, $U,$ and $V$ after cycle 6 to be good (solid black lines), especially considering that the signals are very weak (below 3 \% and 1 \% in circular and 
linear polarization, respectively) and that we only used two nodes for each of the three components of the magnetic field (see Sect.~\ref{sec:inversion}). Of course, cycle 3 produces no polarization signals
(dashed red lines) because the magnetic field was not being inverted at this point.\\ 

\section{Average quet-Sun, granular, intergranular, and dark magnetic element models}

\begin{figure}
\begin{center}
\includegraphics[width=7cm]{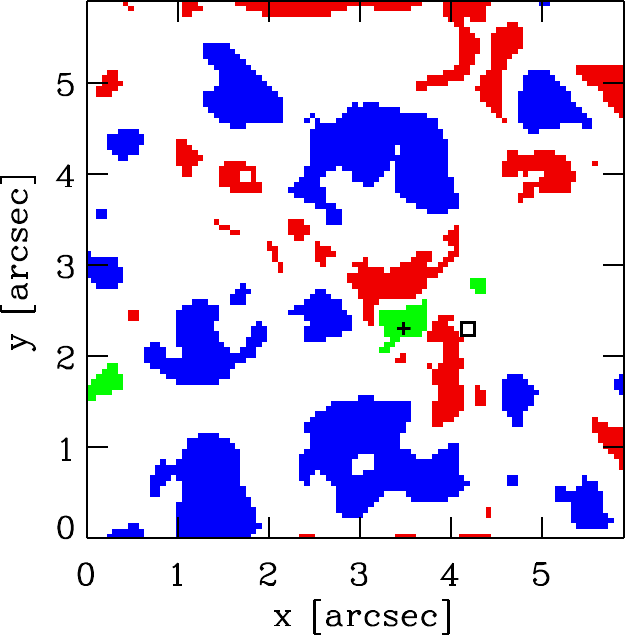}
\end{center}
\caption{Same as Figs.~\ref{fig:continuum},~\ref{fig:phys1}, and~\ref{fig:phys2}, but showing regions ascribed to granules (blue), intergranules (red), and to dark magnetic elements (green).\label{fig:selection}}
\end{figure}

The average quiet-Sun model is determined from all 280,000 pixels, within the 5.9"$\times$5.9" field of view, and all 28 snapshots. For the other three models, we identify pixels as belonging to granules, 
intergranules, or magnetic element depending on two quantities: the vertical component of the velocity and of the magnetic field at an optical depth $\log\tau_{\rm c}=-1$. We refer to these 
two quantities as $v_z^{*}$ and $B_z^{*}$, respectively. On the one hand, granular regions are defined as pixels 
where $v_z^{*}< -1.0$~km~s$^{-1}$ and $\|B_z^{*}\| < 20$~G. On the other hand, intergranular regions are defined as pixels where $v_z^{*} > 0.75$~km~s$^{-1}$ and $\|B_z^{*}\| < 20$~G. In addition,
pixels where $B_z^{*} > 75$~G are considered to correspond to magnetic elements. An example of the selected pixels using these definitions, for the same snapshot as in the previous figures, is displayed
in Figure.~\ref{fig:selection}. We note that the pixels selected as magnetic elements lie mostly in an intergranular region, and therefore it is better to refer to them as { dark} magnetic element. This is
done in order to distinguish them from bright magnetic elements that appear mostly in the network and plage regions. { Incidentally, our selection criteria for the dark magnetic element only includes pixels where the
magnetic field has positive polarity, and therefore the pixel marked with a square ($\square$) in previous figures is not actually counted in the average.} Once all snapshots are considered, we have a total of about 
30,000, 20,000, and 5,000 pixels belonging to granules, intergranules, and dark magnetic elements, respectively.\\

\begin{figure*}
\begin{tabular}{cc}
\includegraphics[width=8cm]{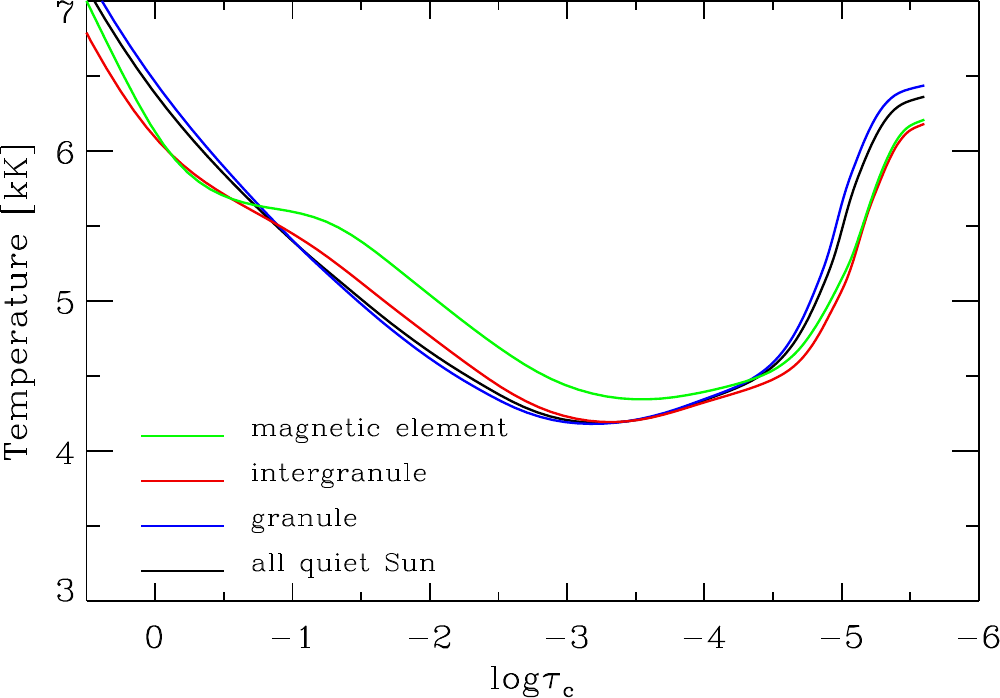} &
\includegraphics[width=8.3cm]{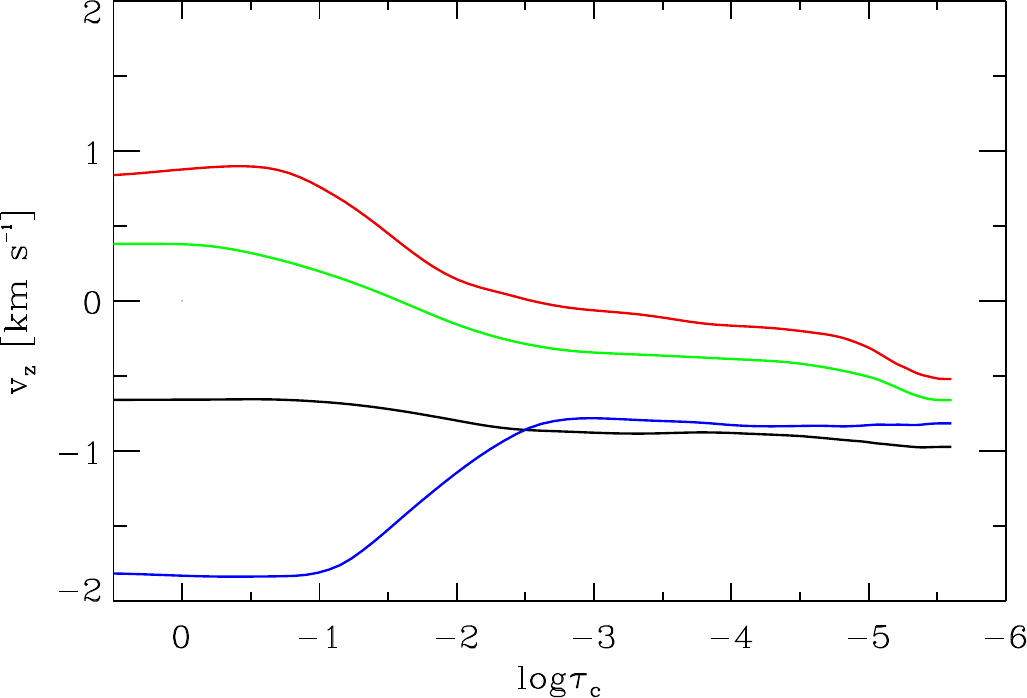} \\
\includegraphics[width=8cm]{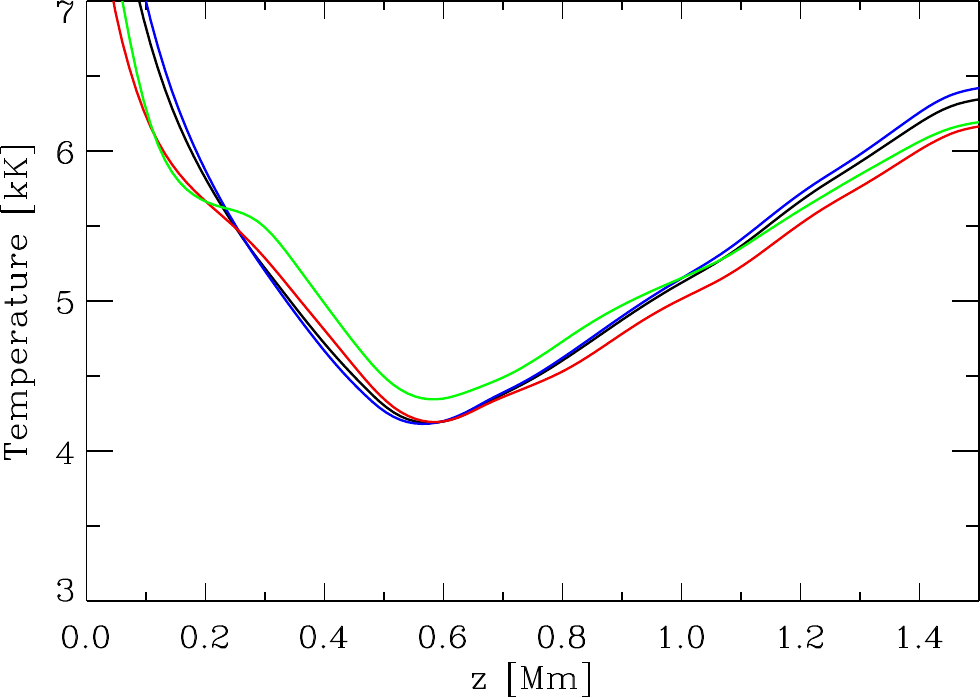} &
\includegraphics[width=8.3cm]{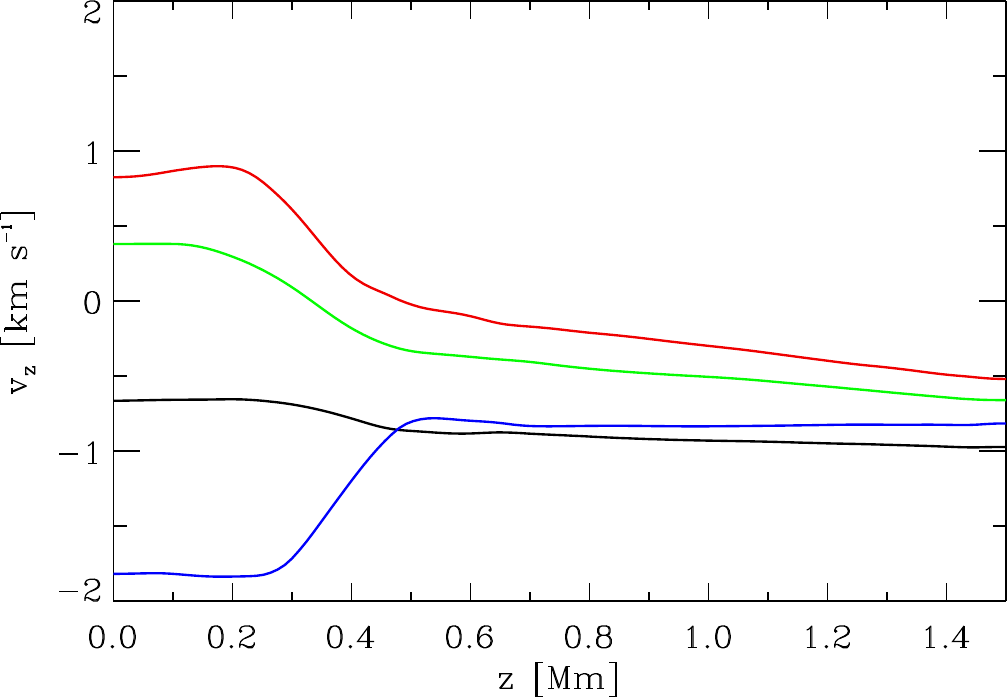}
\end{tabular}
\caption{Optical depth stratifciation ($\log\tau_{\rm c}$; upper panels) and height stratification ($z$; lower panels) of the temperature (left) and vertical velocity (right) in the four one-dimensional 
semi-empirical models presented in this work: average quiet-Sun (black), granule (blue), intergranule (red), dark magnetic element (green).\label{fig:avgmod}}
\end{figure*}

After granular, intergranular, and dark magnetic element pixels are selected, we construct the average (which corresponds to a spatial and temporal average)  in our entire dataset as a function of the
vertical coordinate $z$. This is done for the temperature $T(z)$, gas pressure $P_{\rm g}(z)$, vertical velocity $v_z(z)$, vertical $B_z(z)$, and horizontal $B_h(z)=\sqrt{B_x^2+B_y^2}$ component 
of the magnetic field. The density is not calculated by averaging, but rather via the equation of state from the averaged temperature and gas pressure. Once the density is known, the opacity and
optical depth scales for the spatially and temporarily averaged models can be calculated. The resulting one-dimensional semi-empirical models calculated in this fashion are provided, with a reduced
sampling ($\Delta z=36$~km) in Table~\ref{tab:qs} (quiet Sun), Table~\ref{tab:granule} (granular model), Table~\ref{tab:intergranule} (intergranular model), and Table~\ref{tab:knot} (dark magnetic element model).
Models tabulated with the original $\Delta z=12$~km sampling can be obtained electronically\footnote{\url{https://cdsarc.u-strasbg.fr/ftp/pub/J/A+A/...}}. From all the provided physical parameters, the most 
useful ones are the thermodynamic and kinematic ones. The temperature and vertical velocities are illustrated in Figure~\ref{fig:avgmod} both as a function of $z$ and $\log\tau_{\rm c}$. { Uncertainties
in these models are discussed in Sect.~\ref{sec:errors}. We note that very similar average models would be obtained using only one snapshot from the available time series, but we decided to consider all of them
to account for the different stages in the evolution of the studied features.}\\

It is important to note that comparisons with previously existing semi-empirical models should be done with care, especially if those models were obtained from the inversion of spatially and temporally averaged Stokes 
profiles. As demonstrated by \citet{uitenbroek2011} and \citet{ivan2023}, for example, due to the non-linearity of the radiative transfer equation, the average atmosphere obtained from the inversion 
of many Stokes profiles is not necessarily the same as the atmosphere resulting from the inversion of the average of those Stokes profiles. With this in mind, we now focus on some of the properties of 
the inferred models as presented in this figure.\\

In the average granular (blue) and interganular (red) models, we observe a temperature reversal between granules and intergranules in the $\log\tau_{\rm c} \in [-1.0,-3.0]$ or $z \in [0.25,0.50]$~Mm
region (see left panels in Fig.~\ref{fig:avgmod}). This is in agreement with existing models \citep[][;hereafter referred to as BBR2002]{borrero2002gra}. This reversal gives rise to what is known as 
reverse granulation \citep{gadun1999,cheung2007}. The vertical velocities (right panels in Fig.~\ref{fig:avgmod}) indicate that convective motions are mostly present in the lower photosphere ($\log\tau_{\rm c} \ge -2.0$ 
or $z < 0.40$~Mm). Above this height, intergranular velocities reverse and turn into upflows, while granules are essentially at rest. This is also in quantitative agreement with BBR2002. The main differences between our granular 
and integranular models and those from BBR2002 are that { (a)} we use spatially revolved (with high spatial resolution) observations; { (b)} we extend the models to the lower chromosphere ($\log\tau_{\rm c} \in [-4,-5.6]$); and{ (c)} we also include a more reliable $z$-scale.\\

The temperature in the average dark magnetic element (green) features a large temperature enhancement, with respect to the average quiet-Sun model (black), of about 300-400~K at around $\log\tau_{\rm c}=-2.0$ or $z = 0.4$~Mm.
This is in agreement with network and plage models \citep{solanki1986,solanki1992,lagg2010}. Unlike these aforementioned models, where the temperature enhancement is present at all optical depths, in our model 
the temperature in the deep photosphere is not enhanced with respect to the average quiet-Sun model. This occurs because our magnetic element lies almost entirely in
a dark intergranular lane (see Fig.~\ref{fig:selection}). This also explains why the inferred vertical velocities in the dark magnetic element are so close to that of the intergranules.\\

The $z$ location of the $\log\tau_{\rm c}=0$ (i.e. Wilson depression) can be determined from Tables~\ref{tab:qs},~\ref{tab:granule},~\ref{tab:intergranule}, and ~\ref{tab:knot}. As we can see, the Wilson depression is formed
about 30-40~km higher in granules than in intergranules and the dark magnetic element. We note that, as imposed in Sect.~\ref{sec:inversion}, the values of the gas pressure at the bottom ($z=0$) and top ($z=1524$~km) 
are the same in all three models. We provide a physical parameter in addition to those presented in these tables, which is referred
to as $L_z$ (last column in these tables) and corresponds to the vertical component of the Lorentz force. This one is calculated by applying the $z$ component of the momentum equation:

\begin{equation}
L_z = \frac{1}{c} [\vec{j} \times \vec{B}]_z = \frac{\partial P_{\rm g}}{\partial z} + \rho g
.\end{equation}

\begin{figure*}
\begin{tabular}{cc}
\includegraphics[width=8cm,bb= 75 375 530 710]{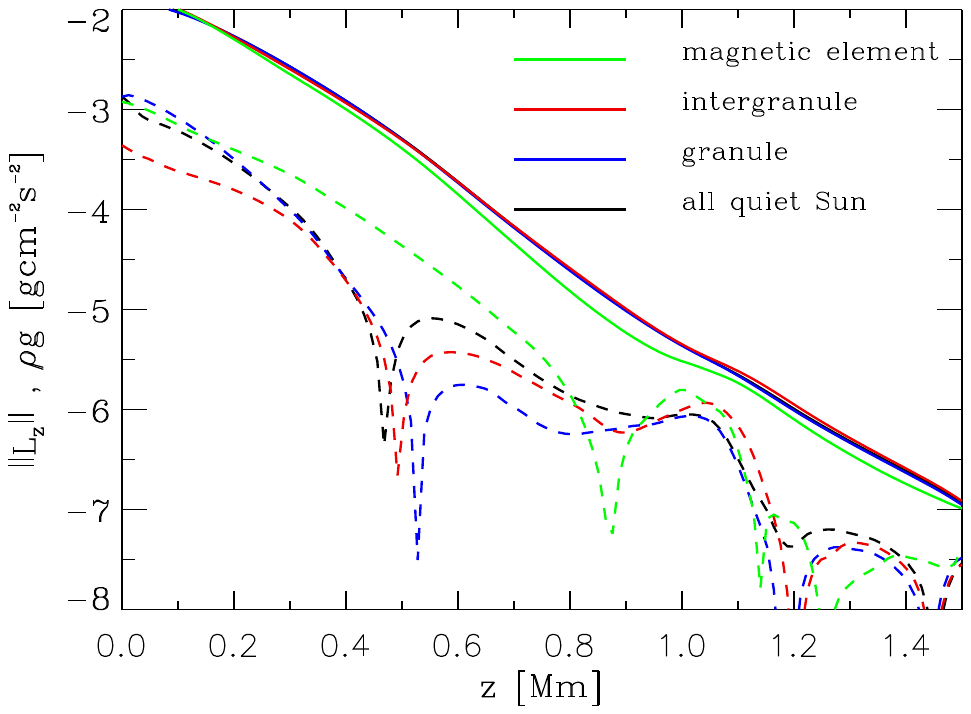} &
\includegraphics[width=8cm,bb= 75 375 530 710]{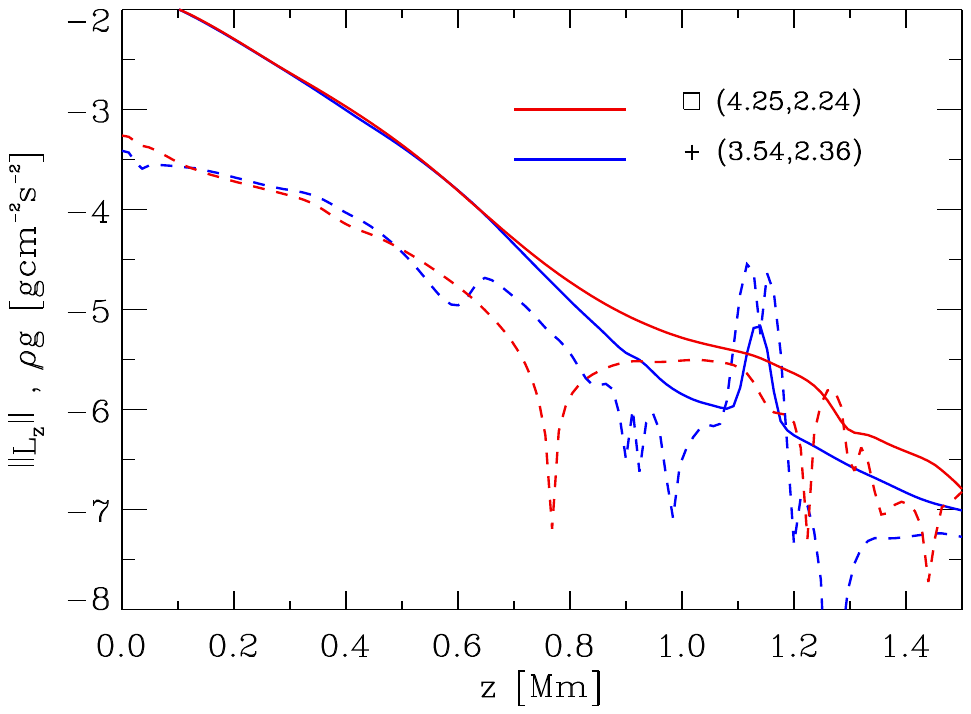}
\end{tabular}
\caption{Comparison between gravity force and vertical component of the Lorentz force. {\it Left panel}: Depth variation of gravity force (solid) and vertical component of the Lorentz force (dashed) in the four one-dimensional semi-empirical models presented in this work: 
average quiet-Sun (black), granule (blue), intergranule (red), dark magnetic element (green). Lorentz force is given in absolute value so that it can be displayed in a vertical logarithmic scale {\it Right panel}: 
Same as above, but considering two individual locations indicated with the $+$ and $\square$ symbols in the previous figures. The approximate $(x,y)$ locations of these two pixels in Figs.~\ref{fig:phys1} and 
~\ref{fig:phys2} are given in parentheses.\label{fig:lorz}}
\end{figure*}

Figure~\ref{fig:lorz} (left panel) presents a comparison of the gravity force $\rho g$ (solid) and the $z$-component of the Lorentz force $L_z$ (dashed) in our four average semi-empirical models. Here, we can see that the gravity force
dominates by one or two orders of magnitude at most heights. Overall, these results indicate that the spatially and temporally averaged models presented here are in quasi-vertical hydrostatic equilibrium. However, at some particular 
depths, the $z$ component of the Lorentz force can become comparable to the gravity force. This is the case of the dark magnetic element (green), where at $z \approx 1.0$~Mm, the Lorentz force amounts to about 30-40 \% of the 
gravity force.\\

This effect is more noticeable when studying the spatially resolved (i.e. non-averaged) results in $(x,y)$. For instance, in the right panel in Fig.~\ref{fig:lorz} we display again the gravity and Lorentz forces, but now focusing on the two pixels indicated with the $+$ and $\square$ symbols in for example Fig.~\ref{fig:phys1},~\ref{fig:phys2}. The region where the Lorenz and gravity forces are comparable ($L_z \sim \rho g$) is now larger. In addition, locally at $z \approx 1.1$~Mm (blue) or at $z \approx 1.25$~Mm (red), the $z$ component of the Lorentz force can actually be larger than the gravity force.
In this regard, these two particular locations selected for Fig.~\ref{fig:lorz} (left panel) are certainly not in vertical hydrostatic equilibrium, with the case of the pixel indicated with $+$ (blue) 
being particularly striking. We note that in the case of spatially resolved results at every $(x,y)$ location on the map, $P_{\rm g}$ is not only affected by the vertical component of the Lorentz force, but also 
by its horizontal ($L_x$ and $L_y$) components because we consider three dimensional magneto-hydrostatic equilibrium \citep[see Eq.~7 in][]{borrero2019mhs}.\\

As shown in Figure~\ref{fig:lorz} (left panel), the effect of the Lorentz force in the determination of the gas pressure in the one-dimensional (spatially averaged) semi-empirical models and, hence, of the 
geometrical height scale $z$ of these models, is minimal. Consequently, there is very little difference between having considered vertical hydrostatic or three-dimensional magneto-hydrostatic equilibrium in the determination of the four average models presented here.\\

Besides the aforementioned discussion about the existence (or non-existence) of quasi-vertical hydrostatic equilibrium in the average and spatially resolved results, there is an additional conceptual difference
between the HE and the MHS approaches. As discussed in \citet{borrero2019mhs} (see Section 3), imposing hydrostatic equilibrium along the three spatial coordinates $(x,y,z)$ leads to a temperature that does not
vary in planes of constant $z$. Since the temperature is a function of $(x,y,z)$, this immediately implies that hydrostatic equilibrium cannot be maintained along all three Cartesian coordinates. So, 
in practice it is only imposed vertically (i.e. along $z$), and the inconsistency along $(x,y)$ is ignored. However, when we account for magneto-hydrostatic equilibrium in three dimensions, we can simultaneously impose vertical 
and horizontal equilibrium with a temperature that also varies as a function of $(x,y),$ even if the Lorentz force plays a small role. In this regard, the MHS approach is preferable to the HE approach.

\section{Uncertainties in the average models}
\label{sec:errors}

\begin{figure*}
\begin{tabular}{cc}
\includegraphics[width=8cm]{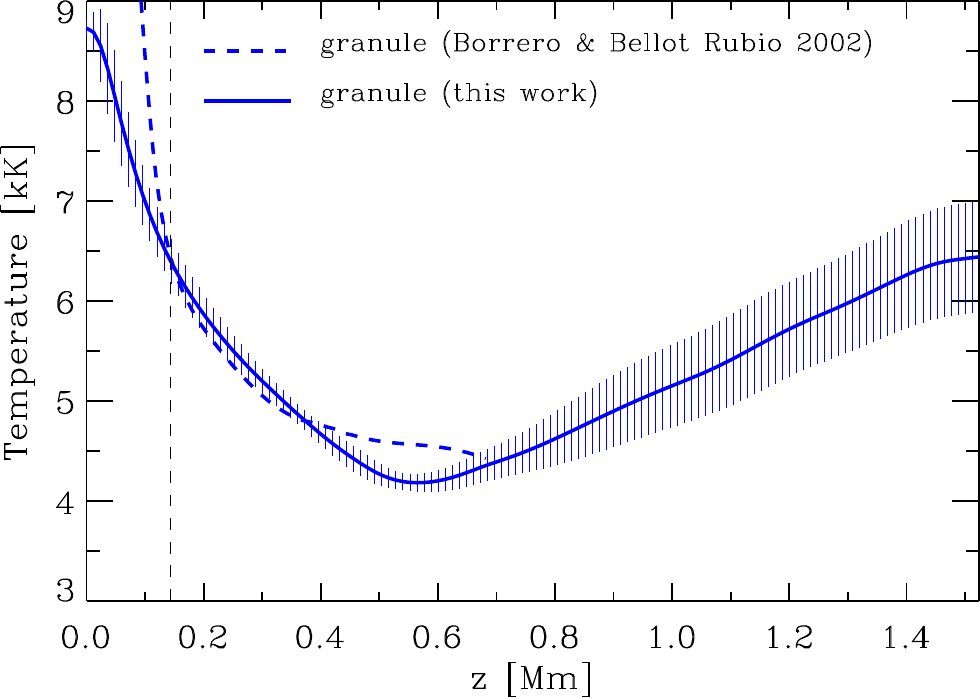} &
\includegraphics[width=8cm]{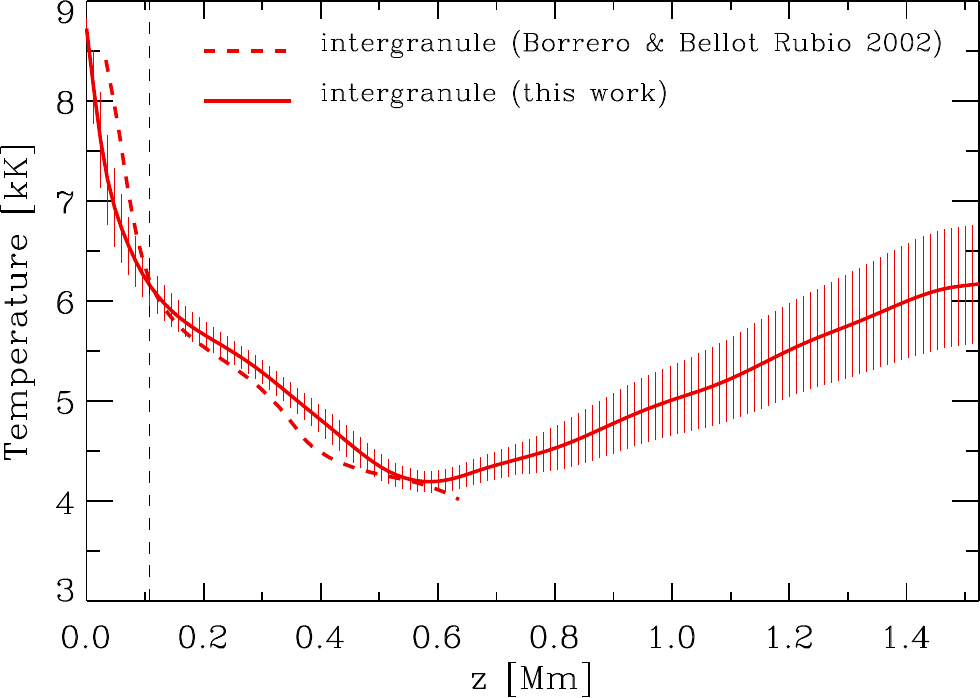} \\
\includegraphics[width=8cm]{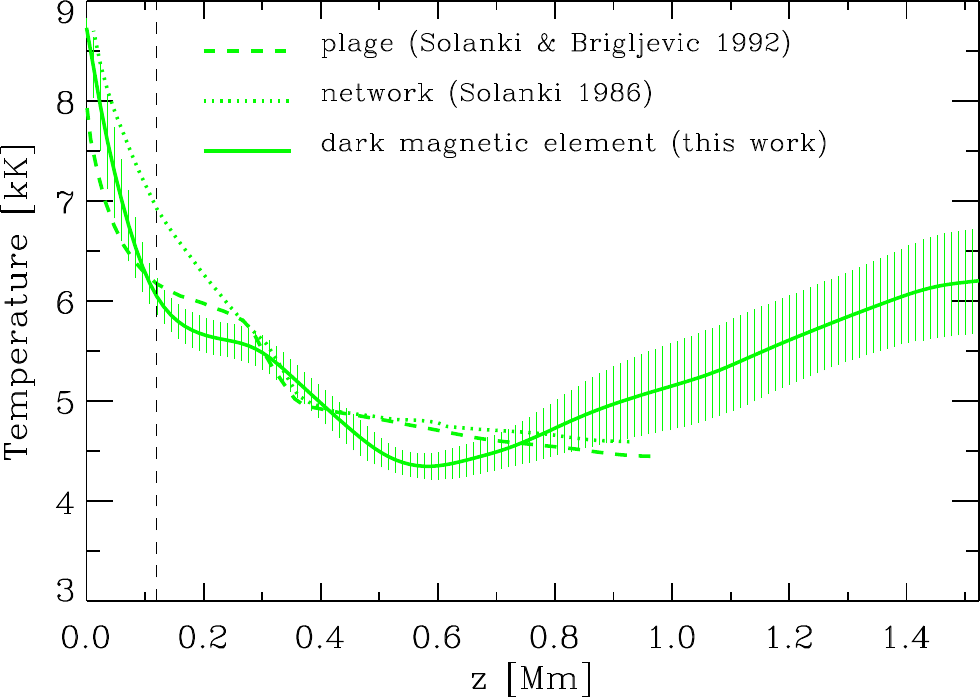} &
\includegraphics[width=8cm]{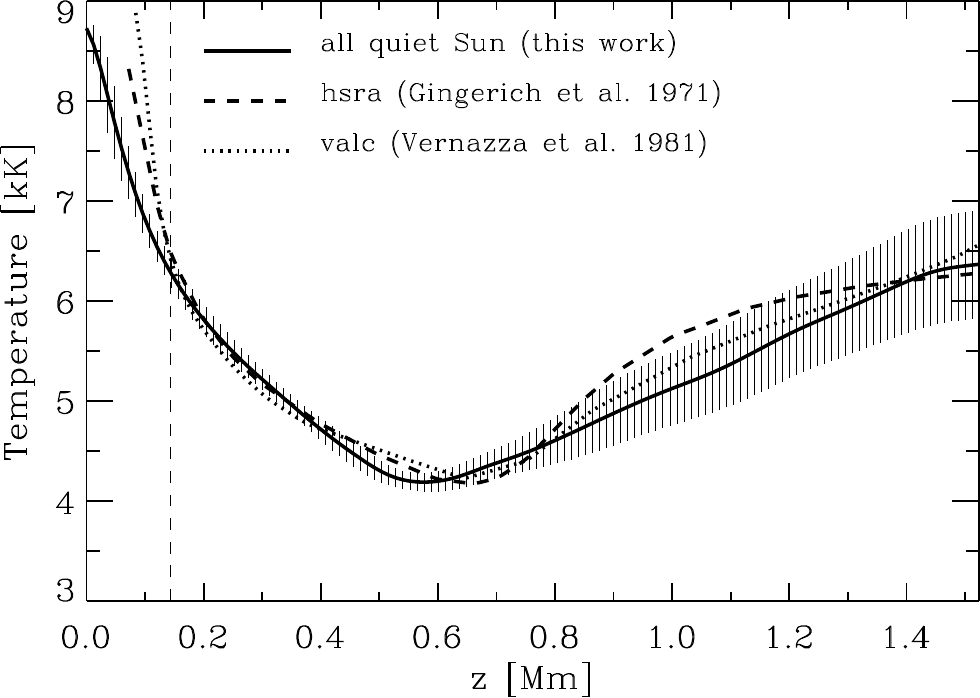}
\end{tabular}
\caption{$T(z)$ corresponding to four models presented in this work: average granule (top left; solid blue), average intergranule (top right; solid red), average dark magnetic elements (bottom left; solid green),
and average quiet Sun (bottom right; solid black). Standard deviation from the mean values are indicated by the vertical colored lines. The vertical black dashed lines show the location of the $\tau_{\rm c}=1$ level in
each of these models. Dashed and dotted color lines represent well-known, one-dimensional semi-empirical models for similar solar structures. The $z$ scale in these models has been shifted so that the $\tau_{\rm c}=1$
coincides with that of the models introduced in this work.\label{fig:compmod}}
\end{figure*}

{ As explained in Sect.~\ref{sec:inversion}, the inversion of the observed Stokes profiles $\vec{I}_{\rm obs}(\lambda)$ at each $(x,y)$ pixel on the observed field of view yields the physical parameters 
(temperature $T$, line-of-sight velocity $v_{\rm z}$, etc.) as a function of $(x,y,z)$. Along with it, the FIRTEZ code also yields an error in the estimation of the physical parameters based on the 
formulation described in Appendix B or Chapter 11.2.1 in \citet{jorge1997} and \citet{jc2003book}, respectively. As a measure of the errors in the determination of the physical parameters of the average
one-dimensional models presented in this work, we could consider the average of the individual errors at each pixel. Instead of this, it is more representative of the actual uncertainties to consider the 
standard deviation, around the mean value, from the ensemble of thousands of pixels used to construct these models. These uncertainties are more conservative (i.e. larger) than the former ones because 
they also include the effects of averaging many pixels that, although they have been ascribed to belong to the same feature (e.g. granule, intergranule, dark magnetic element), they can represent different locations
within them (e.g. edge or center of granules) or different stages in the temporal evolution.\\

Figure~\ref{fig:compmod} presents the average $T(z)$ for each of the models presented: granule (top left; blue lines), intergranule (top right; red lines), dark magnetic elements (bottom left; green lines),
and full quiet Sun (bottom right; black lines). These are essentially the same as those in the bottom left panel in Fig.~\ref{fig:avgmod}, with the exception that we now also include the standard deviations 
(i.e. uncertainties) in those models as defined above. For comparison purposes, we also display the temperature stratifications as a function of the $z$ coordinate, in a number of widely used models
that represent similar features in the solar atmosphere.\\}

\section{Conclusions and future work}

We applied a newly developed FIRTEZ Stokes inversion code in order to invert high spatial resolution spectropolarimetric observations of the quiet Sun in two spectral lines: one photospheric 
(\ion{Fe}{I} 617.3 nm) and one photospheric-chromospheric (\ion{Ca}{II} 854.2 nm). While the
former spectral line is treated under the assumption of local thermodynamic equilibrium, the latter is treated under non-local thermal equilibrium. The inversion is applied to all pixels on the field
of view and for several observing times. Pixels are then selected as belonging to granules ($\approx 30,000$ pixels), intergranules ($\approx 30,000$ pixels), and dark magnetic element ($\approx 4,000$ pixels). 
With this, an average one-dimensional semi-empirical model is constructed for each of these aforementioned structures. In addition, all pixels in the field of view and at all times ($\approx 280,000$ pixels)
are employed to produce an average model of the entire quiet Sun. Our models are tabulated between $z=[0,1.5]$~Mm or $\log\tau_{\rm c} \simeq [0.75,-5.6]$. The vertical scale $z$ is obtained not from the application
of vertical hydrostatic equilibrium, but instead from the more general three-dimensional magneto-hydrostatic equilibrium (i.e. including the Lorentz force in the pressure balance).{ The models presented in this work 
are therefore particularly useful when the geometrical stratification (i.e. $z$) is needed instead of the optical depth stratification ($\tau_{\rm c}$) or when accurate gas pressures or densities are needed}. We find that, while at individual 
locations the Lorentz force can play an important role (i.e. comparable to or larger than the gravity force), the average models are in a quasi-vertical hydrostatic equilibrium. This will likely not be the case in regions 
on the solar surface that harbor strong magnetic fields (i.e. network, umbra, penumbra, and so forth). We will provide average models for those structures in the magnetized solar atmosphere in a future paper.
\begin{acknowledgements}
This research has made use of NASA's Astrophysics Data System. The Atomic Database (ASD) of the National Institute of Standards
and Technology (NIST) was employed in this work. The Institute for Solar Physics is supported by a grant for research infrastructures 
of national importance from the Swedish Research Council (registration number 2021-00169). The Swedish 1-m Solar Telescope is operated 
on the island of La Palma by the Institute for Solar Physics of Stockholm University in the Spanish Observatorio del Roque de los Muchachos of the 
Instituto de Astrof\'isica de Canarias. APY and JdlCR gratefully acknowledge financial support from the European Union through the European Research Council (ERC) 
under the Horizon Europe program (MAGHEAT, grant agreement 101088184). Views and opinions expressed are however those of the author(s) only and do not necessarily 
reflect those of the European Union or the European Research Council. Neither the European Union nor the granting authority can be held responsible for them.
\end{acknowledgements}

\bibliographystyle{aa}
\bibliography{ms}

\clearpage

\onecolumn

\begin{appendix}
\section{Atmospheric models}
\label{sec:models}

\begin{longtable}{ccccccccccc}
\caption{Average quiet-Sun model.\label{tab:qs}}\\
\hline\hline
$z$ & $\log\tau_c$ & $T$ & $\|B_z\|$ & $B_h$ & $v_z$ & $P_{\rm g}$ & $\rho g$ & $L_z$ & n$_{\rm elec}$ & n$_{\rm hyd}$\\
$\textrm{[km]}$ & & [K] & [G] & [G] & [km s$^{-1}$] & [dyn cm$^{-2}$] & [g s$^{-2}$ cm$^{-2}$] & [g s$^{-2}$ cm$^{-2}$] & [cm$^{-3}$] & [cm$^{-3}$]\\
\hline\hline                                                               
\endfirsthead
\caption{Average quiet-Sun model.}\\                                                  
\hline\hline                                                                                                                 
$z$ & $\log\tau_c$ & $T$ & $\|B_z\|$ & $B_h$ & $v_z$ & $P_{\rm g}$ & $\rho g$ & $L_z$ & n$_{\rm elec}$ & n$_{\rm hyd}$ \\
$\textrm{[km]}$ & & [K] & [G] & [G] & [km s$^{-1}$] & [dyn cm$^{-2}$] & [g s$^{-2}$ cm$^{-2}$] & [g s$^{-2}$ cm$^{-2}$] & [cm$^{-3}$] & [cm$^{-3}$]\\
\hline
\endhead
\endfoot
    0 &    1.482 &  8730.17 &     3.87 &    33.95  &   -0.666  &  2.7566e+05  &  1.3455e-02  & -1.3489e-03 &  2.4379e+15 &  2.0836e+17\\
   36 &    1.047 &  8060.43 &     3.87 &    33.95  &   -0.664  &  2.2628e+05  &  1.2029e-02  & -8.5427e-04 &  1.0317e+15 &  1.8628e+17\\
   72 &    0.622 &  7290.25 &     3.87 &    33.95  &   -0.661  &  1.8205e+05  &  1.0735e-02  & -7.0265e-04 &  3.2778e+14 &  1.6625e+17\\
  108 &    0.246 &  6705.00 &     3.87 &    33.95  &   -0.659  &  1.4361e+05  &  9.2171e-03  & -5.8293e-04 &  1.1584e+14 &  1.4275e+17\\
  144 &   -0.083 &  6283.08 &     3.87 &    33.95  &   -0.658  &  1.1145e+05  &  7.6361e-03  & -4.5557e-04 &  4.9170e+13 &  1.1827e+17\\
  180 &   -0.379 &  5965.31 &     3.87 &    33.95  &   -0.656  &  8.5348e+04  &  6.1601e-03  & -3.4447e-04 &  2.4484e+13 &  9.5416e+16\\
  216 &   -0.651 &  5705.94 &     3.87 &    33.95  &   -0.657  &  6.4628e+04  &  4.8770e-03  & -2.5009e-04 &  1.3748e+13 &  7.5545e+16\\
  252 &   -0.906 &  5480.17 &     3.87 &    33.95  &   -0.666  &  4.8439e+04  &  3.8061e-03  & -1.7602e-04 &  8.4572e+12 &  5.8959e+16\\
  288 &   -1.150 &  5279.15 &     3.87 &    33.95  &   -0.683  &  3.5952e+04  &  2.9325e-03  & -1.1996e-04 &  5.5513e+12 &  4.5429e+16\\
  324 &   -1.392 &  5094.04 &     3.87 &    33.90  &   -0.707  &  2.6433e+04  &  2.2345e-03  & -7.8034e-05 &  3.7811e+12 &  3.4616e+16\\
  360 &   -1.636 &  4910.38 &     3.82 &    32.18  &   -0.739  &  1.9250e+04  &  1.6881e-03  & -4.3524e-05 &  2.6175e+12 &  2.6153e+16\\
  396 &   -1.887 &  4734.69 &     3.68 &    26.79  &   -0.779  &  1.3884e+04  &  1.2628e-03  & -2.1411e-05 &  1.8195e+12 &  1.9564e+16\\
  432 &   -2.147 &  4571.69 &     3.53 &    22.39  &   -0.822  &  9.9098e+03  &  9.3344e-04  & -8.9752e-06 &  1.2571e+12 &  1.4463e+16\\
  468 &   -2.415 &  4422.25 &     3.41 &    19.54  &   -0.854  &  7.0005e+03  &  6.8169e-04  & -4.3499e-07 &  8.6088e+11 &  1.0562e+16\\
  504 &   -2.690 &  4290.87 &     3.31 &    17.24  &   -0.868  &  4.8969e+03  &  4.9145e-04  &  5.5331e-06 &  5.8675e+11 &  7.6149e+15\\
  540 &   -2.967 &  4210.69 &     3.24 &    15.38  &   -0.877  &  3.4014e+03  &  3.4786e-04  &  8.0297e-06 &  4.0912e+11 &  5.3897e+15\\
  576 &   -3.244 &  4186.83 &     3.19 &    13.98  &   -0.884  &  2.3572e+03  &  2.4245e-04  &  7.8619e-06 &  2.9522e+11 &  3.7558e+15\\
  612 &   -3.517 &  4207.82 &     3.18 &    13.00  &   -0.882  &  1.6367e+03  &  1.6750e-04  &  6.6239e-06 &  2.1930e+11 &  2.5944e+15\\
  648 &   -3.780 &  4267.09 &     3.18 &    12.36  &   -0.877  &  1.1423e+03  &  1.1528e-04  &  5.1192e-06 &  1.6623e+11 &  1.7853e+15\\
  684 &   -4.022 &  4345.39 &     3.21 &    11.96  &   -0.881  &  8.0321e+02  &  7.9596e-05  &  3.6255e-06 &  1.2649e+11 &  1.2327e+15\\
  720 &   -4.229 &  4417.84 &     3.26 &    11.77  &   -0.889  &  5.6843e+02  &  5.5406e-05  &  2.6060e-06 &  9.6090e+10 &  8.5800e+14\\
  756 &   -4.396 &  4494.72 &     3.35 &    11.71  &   -0.896  &  4.0466e+02  &  3.8768e-05  &  1.9014e-06 &  7.4760e+10 &  6.0034e+14\\
  792 &   -4.523 &  4582.11 &     3.47 &    11.80  &   -0.903  &  2.8996e+02  &  2.7249e-05  &  1.4575e-06 &  6.1963e+10 &  4.2196e+14\\
  828 &   -4.618 &  4677.13 &     3.61 &    11.99  &   -0.910  &  2.0931e+02  &  1.9270e-05  &  1.1866e-06 &  5.6276e+10 &  2.9840e+14\\
  864 &   -4.691 &  4775.16 &     3.75 &    12.15  &   -0.916  &  1.5247e+02  &  1.3748e-05  &  1.0107e-06 &  5.5189e+10 &  2.1289e+14\\
  900 &   -4.751 &  4872.70 &     3.91 &    12.16  &   -0.921  &  1.1224e+02  &  9.9173e-06  &  8.9977e-07 &  5.6242e+10 &  1.5357e+14\\
  936 &   -4.800 &  4967.24 &     4.07 &    11.92  &   -0.925  &  8.3756e+01  &  7.2587e-06  &  8.2872e-07 &  5.7910e+10 &  1.1240e+14\\
  972 &   -4.843 &  5056.26 &     4.23 &    11.34  &   -0.929  &  6.3595e+01  &  5.4134e-06  &  8.4678e-07 &  5.9464e+10 &  8.3828e+13\\
 1008 &   -4.882 &  5139.17 &     4.40 &    10.39  &   -0.932  &  4.9385e+01  &  4.1350e-06  &  8.9457e-07 &  6.0820e+10 &  6.4031e+13\\
 1044 &   -4.916 &  5220.42 &     4.57 &     9.07  &   -0.934  &  3.9315e+01  &  3.2395e-06  &  8.2657e-07 &  6.2294e+10 &  5.0164e+13\\
 1080 &   -4.950 &  5309.49 &     4.75 &     7.61  &   -0.937  &  3.1575e+01  &  2.5568e-06  &  5.1386e-07 &  6.3808e+10 &  3.9592e+13\\
 1116 &   -4.982 &  5411.67 &     4.91 &     6.54  &   -0.940  &  2.4933e+01  &  1.9792e-06  &  2.1541e-07 &  6.4566e+10 &  3.0648e+13\\
 1152 &   -5.015 &  5522.00 &     5.02 &     6.10  &   -0.944  &  1.9283e+01  &  1.4981e-06  &  9.6812e-08 &  6.3961e+10 &  2.3198e+13\\
 1188 &   -5.048 &  5631.30 &     5.07 &     6.02  &   -0.948  &  1.4731e+01  &  1.1199e-06  &  4.2834e-08 &  6.1796e+10 &  1.7342e+13\\
 1224 &   -5.083 &  5732.64 &     5.09 &     6.10  &   -0.952  &  1.1291e+01  &  8.4081e-07  &  5.6649e-08 &  5.8479e+10 &  1.3020e+13\\
 1260 &   -5.119 &  5824.47 &     5.10 &     6.18  &   -0.955  &  8.7784e+00  &  6.4100e-07  &  6.3164e-08 &  5.4619e+10 &  9.9259e+12\\
 1296 &   -5.158 &  5914.82 &     5.11 &     6.23  &   -0.959  &  6.9163e+00  &  4.9481e-07  &  5.8646e-08 &  5.0612e+10 &  7.6621e+12\\
 1332 &   -5.201 &  6008.49 &     5.11 &     6.26  &   -0.963  &  5.5085e+00  &  3.8526e-07  &  5.2382e-08 &  4.6604e+10 &  5.9655e+12\\
 1368 &   -5.251 &  6104.22 &     5.11 &     6.27  &   -0.967  &  4.4214e+00  &  3.0150e-07  &  4.1872e-08 &  4.2494e+10 &  4.6683e+12\\
 1404 &   -5.312 &  6197.39 &     5.11 &     6.27  &   -0.972  &  3.5613e+00  &  2.3623e-07  &  2.8399e-08 &  3.8046e+10 &  3.6575e+12\\
 1440 &   -5.385 &  6275.95 &     5.11 &     6.27  &   -0.976  &  2.8556e+00  &  1.8432e-07  &  9.3573e-09 &  3.2870e+10 &  2.8535e+12\\
 1476 &   -5.471 &  6325.30 &     5.11 &     6.27  &   -0.974  &  2.2344e+00  &  1.4096e-07  & -1.6962e-08 &  2.6465e+10 &  2.1820e+12\\
 1512 &   -5.568 &  6353.00 &     5.11 &     6.27  &   -0.973  &  1.6754e+00  &  1.0354e-07  & -2.1209e-08 &  1.7583e+10 &  1.6027e+12  
\end{longtable}                                                                                                            
\tablefoot{Model tabulated with $\Delta z=36$~km. The full model with $\Delta z=12$~km can be obtained electronically at the CDS (qs.dat).}

\begin{longtable}{ccccccccccc}
\caption{Average granular model.\label{tab:granule}}\\
\hline\hline
$z$ & $\log\tau_c$ & $T$ & $\|B_z\|$ & $B_h$ & $v_z$ & $P_{\rm g}$ & $\rho g$ & $L_z$ & n$_{\rm elec}$ & n$_{\rm hyd}$\\
$\textrm{[km]}$ & & [K] & [G] & [G] & [km s$^{-1}$] & [dyn cm$^{-2}$] & [g s$^{-2}$ cm$^{-2}$] & [g s$^{-2}$ cm$^{-2}$] & [cm$^{-3}$] & [cm$^{-3}$]\\
\hline\hline                                                               
\endfirsthead
\caption{Average granular model.}\\                                                  
\hline\hline                                                                                                                 
$z$ & $\log\tau_c$ & $T$ & $\|B_z\|$ & $B_h$ & $v_z$ & $P_{\rm g}$ & $\rho g$ & $L_z$ & n$_{\rm elec}$ & n$_{\rm hyd}$ \\
$\textrm{[km]}$ & & [K] & [G] & [G] & [km s$^{-1}$] & [dyn cm$^{-2}$] & [g s$^{-2}$ cm$^{-2}$] & [g s$^{-2}$ cm$^{-2}$] & [cm$^{-3}$] & [cm$^{-3}$]\\
\hline
\endhead
\endfoot
    0 &    1.553 &  8728.56 &     3.93 &    39.93  &   -1.820  &  2.7568e+05  &  1.3459e-02  & -1.3452e-03 &  2.4339e+15 &  2.0841e+17\\
   36 &    1.145 &  8329.42 &     3.93 &    39.93  &   -1.818  &  2.2618e+05  &  1.1611e-02  & -1.2726e-03 &  1.4195e+15 &  1.7980e+17\\
   72 &    0.700 &  7516.44 &     3.93 &    39.93  &   -1.815  &  1.8237e+05  &  1.0422e-02  & -1.0153e-03 &  4.5351e+14 &  1.6140e+17\\
  108 &    0.301 &  6867.47 &     3.93 &    39.93  &   -1.822  &  1.4409e+05  &  9.0269e-03  & -7.7315e-04 &  1.5129e+14 &  1.3980e+17\\
  144 &   -0.051 &  6395.80 &     3.93 &    39.93  &   -1.833  &  1.1190e+05  &  7.5311e-03  & -5.6053e-04 &  5.9829e+13 &  1.1664e+17\\
  180 &   -0.364 &  6037.47 &     3.93 &    39.93  &   -1.838  &  8.5694e+04  &  6.1110e-03  & -3.9363e-04 &  2.7676e+13 &  9.4652e+16\\
  216 &   -0.646 &  5745.15 &     3.93 &    39.93  &   -1.837  &  6.4843e+04  &  4.8598e-03  & -2.6731e-04 &  1.4612e+13 &  7.5277e+16\\
  252 &   -0.906 &  5492.33 &     3.93 &    39.93  &   -1.826  &  4.8530e+04  &  3.8048e-03  & -1.7732e-04 &  8.6246e+12 &  5.8938e+16\\
  288 &   -1.152 &  5268.13 &     3.93 &    39.93  &   -1.761  &  3.5941e+04  &  2.9378e-03  & -1.1472e-04 &  5.5278e+12 &  4.5510e+16\\
  324 &   -1.394 &  5064.56 &     3.93 &    39.88  &   -1.607  &  2.6348e+04  &  2.2403e-03  & -7.2250e-05 &  3.7204e+12 &  3.4706e+16\\
  360 &   -1.641 &  4869.41 &     3.87 &    38.14  &   -1.414  &  1.9122e+04  &  1.6910e-03  & -4.0636e-05 &  2.5581e+12 &  2.6199e+16\\
  396 &   -1.895 &  4685.95 &     3.68 &    32.69  &   -1.220  &  1.3738e+04  &  1.2625e-03  & -2.1690e-05 &  1.7635e+12 &  1.9561e+16\\
  432 &   -2.159 &  4518.93 &     3.48 &    27.73  &   -1.039  &  9.7635e+03  &  9.3040e-04  & -1.2015e-05 &  1.2075e+12 &  1.4417e+16\\
  468 &   -2.432 &  4373.24 &     3.32 &    24.17  &   -0.889  &  6.8659e+03  &  6.7608e-04  & -6.0473e-06 &  8.2269e+11 &  1.0476e+16\\
  504 &   -2.709 &  4255.05 &     3.20 &    21.17  &   -0.801  &  4.7838e+03  &  4.8414e-04  & -1.7748e-06 &  5.6220e+11 &  7.5021e+15\\
  540 &   -2.987 &  4191.88 &     3.11 &    18.69  &   -0.782  &  3.3141e+03  &  3.4046e-04  &  6.2279e-07 &  3.9531e+11 &  5.2750e+15\\
  576 &   -3.264 &  4181.95 &     3.05 &    16.76  &   -0.792  &  2.2934e+03  &  2.3616e-04  &  1.5751e-06 &  2.8761e+11 &  3.6584e+15\\
  612 &   -3.537 &  4213.84 &     3.03 &    15.44  &   -0.802  &  1.5915e+03  &  1.6264e-04  &  1.7653e-06 &  2.1502e+11 &  2.5191e+15\\
  648 &   -3.798 &  4278.93 &     3.03 &    14.58  &   -0.813  &  1.1107e+03  &  1.1178e-04  &  1.6200e-06 &  1.6339e+11 &  1.7311e+15\\
  684 &   -4.035 &  4356.84 &     3.06 &    14.12  &   -0.830  &  7.8070e+02  &  7.7162e-05  &  1.1915e-06 &  1.2425e+11 &  1.1950e+15\\
  720 &   -4.235 &  4428.27 &     3.11 &    13.88  &   -0.836  &  5.5215e+02  &  5.3692e-05  &  8.9241e-07 &  9.4555e+10 &  8.3146e+14\\
  756 &   -4.393 &  4507.91 &     3.21 &    13.89  &   -0.835  &  3.9286e+02  &  3.7527e-05  &  6.6077e-07 &  7.4232e+10 &  5.8113e+14\\
  792 &   -4.511 &  4598.68 &     3.34 &    13.94  &   -0.834  &  2.8155e+02  &  2.6363e-05  &  5.7180e-07 &  6.2652e+10 &  4.0824e+14\\
  828 &   -4.599 &  4696.64 &     3.49 &    13.93  &   -0.833  &  2.0359e+02  &  1.8665e-05  &  5.8203e-07 &  5.8184e+10 &  2.8904e+14\\
  864 &   -4.667 &  4797.21 &     3.67 &    13.75  &   -0.833  &  1.4883e+02  &  1.3358e-05  &  6.2064e-07 &  5.8134e+10 &  2.0685e+14\\
  900 &   -4.723 &  4896.57 &     3.85 &    13.33  &   -0.834  &  1.1015e+02  &  9.6850e-06  &  6.6746e-07 &  6.0047e+10 &  1.4997e+14\\
  936 &   -4.770 &  4992.53 &     4.04 &    12.68  &   -0.836  &  8.2696e+01  &  7.1303e-06  &  7.0032e-07 &  6.2477e+10 &  1.1041e+14\\
  972 &   -4.811 &  5083.39 &     4.22 &    11.72  &   -0.836  &  6.3100e+01  &  5.3424e-06  &  7.7571e-07 &  6.4731e+10 &  8.2728e+13\\
 1008 &   -4.847 &  5169.31 &     4.42 &    10.47  &   -0.836  &  4.9245e+01  &  4.0989e-06  &  8.5849e-07 &  6.6845e+10 &  6.3473e+13\\
 1044 &   -4.881 &  5255.15 &     4.61 &     9.01  &   -0.835  &  3.9241e+01  &  3.2116e-06  &  7.9870e-07 &  6.9047e+10 &  4.9733e+13\\
 1080 &   -4.914 &  5350.03 &     4.80 &     7.64  &   -0.833  &  3.1353e+01  &  2.5190e-06  &  4.7611e-07 &  7.1131e+10 &  3.9008e+13\\
 1116 &   -4.946 &  5457.74 &     4.98 &     6.79  &   -0.832  &  2.4457e+01  &  1.9243e-06  &  1.6053e-07 &  7.2033e+10 &  2.9798e+13\\
 1152 &   -4.978 &  5571.12 &     5.11 &     6.57  &   -0.830  &  1.8758e+01  &  1.4435e-06  &  4.2297e-08 &  7.1355e+10 &  2.2354e+13\\
 1188 &   -5.011 &  5681.01 &     5.18 &     6.60  &   -0.828  &  1.4299e+01  &  1.0766e-06  & -5.1364e-10 &  6.9015e+10 &  1.6671e+13\\
 1224 &   -5.045 &  5781.37 &     5.22 &     6.68  &   -0.825  &  1.0971e+01  &  8.0905e-07  &  2.4882e-08 &  6.5446e+10 &  1.2528e+13\\
 1260 &   -5.081 &  5872.82 &     5.24 &     6.72  &   -0.824  &  8.5480e+00  &  6.1794e-07  &  4.0106e-08 &  6.1336e+10 &  9.5688e+12\\
 1296 &   -5.120 &  5965.45 &     5.24 &     6.73  &   -0.825  &  6.7423e+00  &  4.7707e-07  &  4.0903e-08 &  5.7101e+10 &  7.3873e+12\\
 1332 &   -5.164 &  6064.68 &     5.25 &     6.74  &   -0.825  &  5.3692e+00  &  3.7064e-07  &  3.7763e-08 &  5.2872e+10 &  5.7390e+12\\
 1368 &   -5.216 &  6167.28 &     5.25 &     6.74  &   -0.825  &  4.3124e+00  &  2.8942e-07  &  2.9798e-08 &  4.8520e+10 &  4.4812e+12\\
 1404 &   -5.279 &  6266.89 &     5.25 &     6.74  &   -0.826  &  3.4777e+00  &  2.2628e-07  &  1.8447e-08 &  4.3696e+10 &  3.5033e+12\\
 1440 &   -5.358 &  6349.92 &     5.25 &     6.74  &   -0.827  &  2.7958e+00  &  1.7642e-07  &  1.4594e-09 &  3.7925e+10 &  2.7310e+12\\
 1476 &   -5.453 &  6400.53 &     5.25 &     6.74  &   -0.819  &  2.1990e+00  &  1.3529e-07  & -2.2629e-08 &  3.0647e+10 &  2.0941e+12\\
 1512 &   -5.562 &  6428.15 &     5.25 &     6.74  &   -0.816  &  1.6663e+00  &  1.0024e-07  & -2.4505e-08 &  2.0449e+10 &  1.5515e+12
\end{longtable}                                                                                                            
\tablefoot{Model tabulated with $\Delta z=36$~km. The full model with $\Delta z=12$~km can be obtained electronically at the CDS (granule.dat).}


\begin{longtable}{ccccccccccc}
\caption{Average intergranular model.\label{tab:intergranule}}\\
\hline\hline
$z$ & $\log\tau_c$ & $T$ & $\|B_z\|$ & $B_h$ & $v_z$ & $P_{\rm g}$ & $\rho g$ & $L_z$ & n$_{\rm elec}$ & n$_{\rm hyd}$\\
$\textrm{[km]}$ & & [K] & [G] & [G] & [km s$^{-1}$] & [dyn cm$^{-2}$] & [g s$^{-2}$ cm$^{-2}$] & [g s$^{-2}$ cm$^{-2}$] & [cm$^{-3}$] & [cm$^{-3}$]\\
\hline\hline                                                               
\endfirsthead
\caption{Average intergranular model.}\\                                                  
\hline\hline                                                                                                                 
$z$ & $\log\tau_c$ & $T$ & $\|B_z\|$ & $B_h$ & $v_z$ & $P_{\rm g}$ & $\rho g$ & $L_z$ & n$_{\rm elec}$ & n$_{\rm hyd}$\\
$\textrm{[km]}$ & & [K] & [G] & [G] & [km s$^{-1}$] & [dyn cm$^{-2}$] & [g s$^{-2}$ cm$^{-2}$] & [g s$^{-2}$ cm$^{-2}$] & [cm$^{-3}$] & [cm$^{-3}$]\\
\hline
\endhead
\endfoot
    0 &    1.262 &  8728.83 &     6.24 &    17.92  &    0.824  &  2.7566e+05  &  1.3458e-02  & -4.3829e-04 &  2.4345e+15 &  2.0839e+17\\
   36 &    0.716 &  7218.08 &     6.24 &    17.92  &    0.830  &  2.2527e+05  &  1.3421e-02  & -3.3737e-04 &  3.2903e+14 &  2.0785e+17\\
   72 &    0.356 &  6552.60 &     6.24 &    17.92  &    0.848  &  1.7859e+05  &  1.1731e-02  & -2.7688e-04 &  1.0137e+14 &  1.8171e+17\\
  108 &    0.063 &  6157.44 &     6.24 &    17.92  &    0.872  &  1.3904e+05  &  9.7219e-03  & -2.3646e-04 &  4.5418e+13 &  1.5060e+17\\
  144 &   -0.199 &  5908.54 &     6.24 &    17.92  &    0.890  &  1.0681e+05  &  7.7835e-03  & -2.0438e-04 &  2.5850e+13 &  1.2057e+17\\
  180 &   -0.448 &  5740.36 &     6.24 &    17.92  &    0.898  &  8.1259e+04  &  6.0953e-03  & -1.7349e-04 &  1.6764e+13 &  9.4422e+16\\
  216 &   -0.694 &  5608.28 &     6.24 &    17.92  &    0.874  &  6.1349e+04  &  4.7103e-03  & -1.4389e-04 &  1.1489e+13 &  7.2965e+16\\
  252 &   -0.941 &  5481.01 &     6.24 &    17.92  &    0.789  &  4.5998e+04  &  3.6137e-03  & -1.1504e-04 &  7.9719e+12 &  5.5978e+16\\
  288 &   -1.186 &  5338.28 &     6.24 &    17.92  &    0.660  &  3.4239e+04  &  2.7619e-03  & -8.7315e-05 &  5.4923e+12 &  4.2782e+16\\
  324 &   -1.429 &  5173.19 &     6.24 &    17.85  &    0.501  &  2.5276e+04  &  2.1040e-03  & -6.0630e-05 &  3.7640e+12 &  3.2591e+16\\
  360 &   -1.670 &  4998.18 &     6.16 &    16.20  &    0.329  &  1.8494e+04  &  1.5933e-03  & -3.6656e-05 &  2.6103e+12 &  2.4682e+16\\
  396 &   -1.916 &  4825.05 &     5.92 &    12.01  &    0.182  &  1.3408e+04  &  1.1966e-03  & -2.0881e-05 &  1.8273e+12 &  1.8537e+16\\
  432 &   -2.169 &  4650.70 &     5.65 &     9.71  &    0.090  &  9.6253e+03  &  8.9124e-04  & -1.0181e-05 &  1.2700e+12 &  1.3807e+16\\
  468 &   -2.430 &  4478.66 &     5.41 &     8.96  &    0.028  &  6.8375e+03  &  6.5743e-04  & -3.2380e-06 &  8.6905e+11 &  1.0185e+16\\
  504 &   -2.701 &  4331.01 &     5.19 &     8.47  &   -0.029  &  4.8077e+03  &  4.7802e-04  &  8.9951e-07 &  5.9157e+11 &  7.4062e+15\\
  540 &   -2.975 &  4234.06 &     5.00 &     8.34  &   -0.061  &  3.3549e+03  &  3.4121e-04  &  3.0462e-06 &  4.1020e+11 &  5.2864e+15\\
  576 &   -3.249 &  4193.81 &     4.83 &     8.37  &   -0.083  &  2.3337e+03  &  2.3963e-04  &  3.7118e-06 &  2.9369e+11 &  3.7121e+15\\
  612 &   -3.522 &  4206.67 &     4.68 &     8.58  &   -0.114  &  1.6254e+03  &  1.6639e-04  &  3.6379e-06 &  2.1745e+11 &  2.5772e+15\\
  648 &   -3.786 &  4262.82 &     4.55 &     8.84  &   -0.149  &  1.1383e+03  &  1.1499e-04  &  3.1973e-06 &  1.6481e+11 &  1.7809e+15\\
  684 &   -4.031 &  4332.45 &     4.45 &     9.18  &   -0.167  &  8.0317e+02  &  7.9830e-05  &  2.6260e-06 &  1.2498e+11 &  1.2363e+15\\
  720 &   -4.249 &  4391.62 &     4.37 &     9.53  &   -0.178  &  5.7041e+02  &  5.5931e-05  &  2.0269e-06 &  9.4149e+10 &  8.6614e+14\\
  756 &   -4.431 &  4449.46 &     4.32 &     9.97  &   -0.193  &  4.0742e+02  &  3.9430e-05  &  1.5714e-06 &  7.1594e+10 &  6.1059e+14\\
  792 &   -4.575 &  4513.66 &     4.31 &    10.42  &   -0.209  &  2.9268e+02  &  2.7922e-05  &  1.2305e-06 &  5.6380e+10 &  4.3238e+14\\
  828 &   -4.685 &  4592.07 &     4.33 &    10.98  &   -0.222  &  2.1164e+02  &  1.9846e-05  &  9.3741e-07 &  4.7842e+10 &  3.0731e+14\\
  864 &   -4.770 &  4684.04 &     4.38 &    11.69  &   -0.237  &  1.5402e+02  &  1.4159e-05  &  6.6695e-07 &  4.4371e+10 &  2.1925e+14\\
  900 &   -4.837 &  4780.12 &     4.45 &    12.45  &   -0.253  &  1.1293e+02  &  1.0172e-05  &  5.8875e-07 &  4.3503e+10 &  1.5752e+14\\
  936 &   -4.892 &  4872.26 &     4.54 &    13.01  &   -0.271  &  8.3904e+01  &  7.4140e-06  &  6.8525e-07 &  4.3486e+10 &  1.1481e+14\\
  972 &   -4.938 &  4955.31 &     4.62 &    13.16  &   -0.287  &  6.3639e+01  &  5.5284e-06  &  8.5462e-07 &  4.3541e+10 &  8.5608e+13\\
 1008 &   -4.979 &  5028.68 &     4.71 &    12.72  &   -0.303  &  4.9626e+01  &  4.2474e-06  &  1.0379e-06 &  4.3584e+10 &  6.5773e+13\\
 1044 &   -5.016 &  5098.81 &     4.81 &    11.61  &   -0.319  &  4.0068e+01  &  3.3815e-06  &  1.1690e-06 &  4.3987e+10 &  5.2363e+13\\
 1080 &   -5.051 &  5177.11 &     4.92 &     9.80  &   -0.336  &  3.3005e+01  &  2.7424e-06  &  9.2495e-07 &  4.4779e+10 &  4.2468e+13\\
 1116 &   -5.085 &  5270.56 &     5.00 &     7.87  &   -0.355  &  2.6712e+01  &  2.1791e-06  &  4.8450e-07 &  4.5182e+10 &  3.3744e+13\\
 1152 &   -5.119 &  5374.72 &     5.05 &     6.76  &   -0.374  &  2.0833e+01  &  1.6652e-06  &  1.3598e-07 &  4.4407e+10 &  2.5786e+13\\
 1188 &   -5.153 &  5480.38 &     5.06 &     6.65  &   -0.393  &  1.5870e+01  &  1.2424e-06  &  1.4868e-08 &  4.2387e+10 &  1.9239e+13\\
 1224 &   -5.187 &  5579.28 &     5.05 &     6.87  &   -0.411  &  1.2027e+01  &  9.2308e-07  &  1.5584e-08 &  3.9451e+10 &  1.4294e+13\\
 1260 &   -5.222 &  5667.25 &     5.03 &     7.03  &   -0.427  &  9.2410e+00  &  6.9652e-07  &  3.3432e-08 &  3.6182e+10 &  1.0786e+13\\
 1296 &   -5.258 &  5749.56 &     5.03 &     7.12  &   -0.442  &  7.1917e+00  &  5.3256e-07  &  4.6225e-08 &  3.2860e+10 &  8.2468e+12\\
 1332 &   -5.296 &  5835.61 &     5.03 &     7.16  &   -0.458  &  5.6710e+00  &  4.1189e-07  &  4.4475e-08 &  2.9725e+10 &  6.3781e+12\\
 1368 &   -5.339 &  5925.83 &     5.03 &     7.17  &   -0.477  &  4.5100e+00  &  3.2056e-07  &  3.6599e-08 &  2.6714e+10 &  4.9637e+12\\
 1404 &   -5.388 &  6015.33 &     5.03 &     7.17  &   -0.493  &  3.6005e+00  &  2.4999e-07  &  2.4444e-08 &  2.3669e+10 &  3.8708e+12\\
 1440 &   -5.445 &  6091.87 &     5.03 &     7.17  &   -0.506  &  2.8633e+00  &  1.9431e-07  &  6.3102e-09 &  2.0335e+10 &  3.0085e+12\\
 1476 &   -5.510 &  6143.85 &     5.03 &     7.17  &   -0.518  &  2.2280e+00  &  1.4828e-07  & -1.8259e-08 &  1.6403e+10 &  2.2957e+12\\
 1512 &   -5.579 &  6171.87 &     5.03 &     7.17  &   -0.520  &  1.6725e+00  &  1.0955e-07  & -2.0162e-08 &  1.1038e+10 &  1.6959e+12
\end{longtable}
\tablefoot{Model tabulated with $\Delta z=36$~km. The full model with $\Delta z=12$~km can be obtained electronically at the CDS (intergr.dat).}


\begin{longtable}{ccccccccccc}
\caption{Average dark magnetic element model.\label{tab:knot}}\\
\hline\hline
$z$ & $\log\tau_c$ & $T$ & $\|B_z\|$ & $B_h$ & $v_z$ & $P_{\rm g}$ & $\rho g$ & $L_z$ & n$_{\rm elec}$ & n$_{\rm hyd}$\\
$\textrm{[km]}$ & & [K] & [G] & [G] & [km s$^{-1}$] & [dyn cm$^{-2}$] & [g s$^{-2}$ cm$^{-2}$] & [g s$^{-2}$ cm$^{-2}$] & [cm$^{-3}$] & [cm$^{-3}$]\\
\hline\hline                                                               
\endfirsthead
\caption{Average dark magnetic element model.}\\
\hline\hline                                                                                                                 
$z$ & $\log\tau_c$ & $T$ & $\|B_z\|$ & $B_h$ & $v_z$ & $P_{\rm g}$ & $\rho g$ & $L_z$ & n$_{\rm elec}$ & n$_{\rm hyd}$\\
$\textrm{[km]}$ & & [K] & [G] & [G] & [km s$^{-1}$] & [dyn cm$^{-2}$] & [g s$^{-2}$ cm$^{-2}$] & [g s$^{-2}$ cm$^{-2}$] & [cm$^{-3}$] & [cm$^{-3}$]\\
\hline
\endhead
\endfoot
    0 &    1.340 &  8729.47 &   135.80 &    15.88  &    0.379  &  2.7567e+05  &  1.3457e-02  & -1.1931e-03 &  2.4362e+15 &  2.0838e+17\\
   36 &    0.809 &  7594.04 &   135.80 &    15.88  &    0.380  &  2.2437e+05  &  1.2691e-02  & -1.0383e-03 &  5.6141e+14 &  1.9654e+17\\
   72 &    0.369 &  6754.29 &   135.80 &    15.88  &    0.380  &  1.7761e+05  &  1.1316e-02  & -8.4636e-04 &  1.4080e+14 &  1.7527e+17\\
  108 &    0.033 &  6173.14 &   135.80 &    15.88  &    0.379  &  1.3719e+05  &  9.5680e-03  & -6.6947e-04 &  4.6327e+13 &  1.4821e+17\\
  144 &   -0.236 &  5856.51 &   135.80 &    15.88  &    0.362  &  1.0405e+05  &  7.6499e-03  & -5.3928e-04 &  2.3658e+13 &  1.1851e+17\\
  180 &   -0.482 &  5707.46 &   135.80 &    15.88  &    0.323  &  7.8033e+04  &  5.8871e-03  & -4.4086e-04 &  1.5686e+13 &  9.1197e+16\\
  216 &   -0.726 &  5638.19 &   135.80 &    15.88  &    0.270  &  5.8126e+04  &  4.4391e-03  & -3.6340e-04 &  1.1519e+13 &  6.8762e+16\\
  252 &   -0.978 &  5598.39 &   135.80 &    15.88  &    0.204  &  4.3094e+04  &  3.3145e-03  & -2.9885e-04 &  8.7344e+12 &  5.1338e+16\\
  288 &   -1.244 &  5528.94 &   135.80 &    15.88  &    0.122  &  3.1794e+04  &  2.4761e-03  & -2.4182e-04 &  6.3544e+12 &  3.8351e+16\\
  324 &   -1.516 &  5386.59 &   135.57 &    15.72  &    0.025  &  2.3279e+04  &  1.8609e-03  & -1.8994e-04 &  4.2298e+12 &  2.8822e+16\\
  360 &   -1.784 &  5198.08 &   132.68 &    13.43  &   -0.078  &  1.6864e+04  &  1.3970e-03  & -1.4257e-04 &  2.7310e+12 &  2.1637e+16\\
  396 &   -2.049 &  5004.74 &   126.29 &     9.29  &   -0.172  &  1.2070e+04  &  1.0385e-03  & -1.0648e-04 &  1.8154e+12 &  1.6085e+16\\
  432 &   -2.316 &  4813.68 &   119.68 &     8.00  &   -0.248  &  8.5236e+03  &  7.6250e-04  & -7.9171e-05 &  1.2282e+12 &  1.1810e+16\\
  468 &   -2.594 &  4632.72 &   113.62 &     8.00  &   -0.303  &  5.9348e+03  &  5.5165e-04  & -5.7765e-05 &  8.2934e+11 &  8.5448e+15\\
  504 &   -2.883 &  4480.00 &   107.74 &     7.87  &   -0.337  &  4.0755e+03  &  3.9174e-04  & -4.1842e-05 &  5.5769e+11 &  6.0679e+15\\
  540 &   -3.183 &  4382.56 &   101.97 &     8.02  &   -0.352  &  2.7662e+03  &  2.7180e-04  & -3.0114e-05 &  3.7969e+11 &  4.2100e+15\\
  576 &   -3.486 &  4344.41 &    96.35 &     8.19  &   -0.364  &  1.8634e+03  &  1.8470e-04  & -2.1464e-05 &  2.6429e+11 &  2.8607e+15\\
  612 &   -3.783 &  4358.65 &    90.93 &     8.26  &   -0.377  &  1.2515e+03  &  1.2364e-04  & -1.5148e-05 &  1.8772e+11 &  1.9149e+15\\
  648 &   -4.058 &  4405.38 &    85.72 &     8.36  &   -0.390  &  8.4150e+02  &  8.2255e-05  & -1.0437e-05 &  1.3459e+11 &  1.2738e+15\\
  684 &   -4.294 &  4461.97 &    80.42 &     8.82  &   -0.401  &  5.6791e+02  &  5.4808e-05  & -7.0714e-06 &  9.7218e+10 &  8.4873e+14\\
  720 &   -4.481 &  4528.54 &    74.66 &     9.81  &   -0.416  &  3.8490e+02  &  3.6600e-05  & -4.8547e-06 &  7.2358e+10 &  5.6676e+14\\
  756 &   -4.618 &  4613.40 &    68.66 &    11.09  &   -0.434  &  2.6255e+02  &  2.4506e-05  & -3.0942e-06 &  5.8166e+10 &  3.7948e+14\\
  792 &   -4.718 &  4707.36 &    62.72 &    12.38  &   -0.449  &  1.8141e+02  &  1.6594e-05  & -1.6728e-06 &  5.1710e+10 &  2.5696e+14\\
  828 &   -4.793 &  4803.87 &    57.11 &    13.29  &   -0.462  &  1.2772e+02  &  1.1447e-05  & -6.4430e-07 &  4.9347e+10 &  1.7726e+14\\
  864 &   -4.852 &  4892.10 &    52.05 &    13.92  &   -0.473  &  9.1918e+01  &  8.0892e-06  & -7.7934e-08 &  4.8248e+10 &  1.2526e+14\\
  900 &   -4.900 &  4970.33 &    47.39 &    14.03  &   -0.483  &  6.7665e+01  &  5.8603e-06  &  4.0697e-07 &  4.7246e+10 &  9.0748e+13\\
  936 &   -4.941 &  5040.09 &    43.23 &    13.88  &   -0.492  &  5.1678e+01  &  4.4130e-06  &  8.1511e-07 &  4.6421e+10 &  6.8337e+13\\
  972 &   -4.977 &  5104.18 &    39.59 &    13.41  &   -0.500  &  4.1385e+01  &  3.4890e-06  &  1.2927e-06 &  4.6096e+10 &  5.4027e+13\\
 1008 &   -5.010 &  5165.89 &    36.23 &    12.70  &   -0.508  &  3.5324e+01  &  2.9417e-06  &  1.5601e-06 &  4.6908e+10 &  4.5553e+13\\
 1044 &   -5.044 &  5230.53 &    32.96 &    11.74  &   -0.517  &  3.0401e+01  &  2.4997e-06  &  1.1497e-06 &  4.7793e+10 &  3.8708e+13\\
 1080 &   -5.078 &  5304.76 &    29.77 &    10.11  &   -0.528  &  2.5737e+01  &  2.0856e-06  &  7.8809e-07 &  4.8269e+10 &  3.2296e+13\\
 1116 &   -5.113 &  5393.38 &    26.77 &     8.59  &   -0.541  &  2.0632e+01  &  1.6431e-06  &  1.8814e-07 &  4.7308e+10 &  2.5445e+13\\
 1152 &   -5.147 &  5485.87 &    24.24 &     8.04  &   -0.554  &  1.5631e+01  &  1.2223e-06  & -8.1007e-08 &  4.4428e+10 &  1.8928e+13\\
 1188 &   -5.179 &  5576.98 &    22.29 &     7.85  &   -0.567  &  1.1566e+01  &  8.8800e-07  & -7.6998e-08 &  4.0454e+10 &  1.3751e+13\\
 1224 &   -5.211 &  5664.40 &    20.86 &     7.86  &   -0.580  &  8.5934e+00  &  6.4788e-07  & -4.0579e-08 &  3.6275e+10 &  1.0033e+13\\
 1260 &   -5.244 &  5749.66 &    19.83 &     7.90  &   -0.593  &  6.5261e+00  &  4.8299e-07  &  1.4485e-09 &  3.2455e+10 &  7.4792e+12\\
 1296 &   -5.277 &  5833.11 &    19.20 &     7.94  &   -0.606  &  5.0531e+00  &  3.6687e-07  &  1.6994e-08 &  2.9003e+10 &  5.6809e+12\\
 1332 &   -5.314 &  5913.68 &    18.84 &     7.91  &   -0.620  &  3.9682e+00  &  2.8241e-07  &  2.3665e-08 &  2.5834e+10 &  4.3730e+12\\
 1368 &   -5.354 &  5993.93 &    18.60 &     7.90  &   -0.633  &  3.1691e+00  &  2.2074e-07  &  3.2444e-08 &  2.2981e+10 &  3.4178e+12\\
 1404 &   -5.400 &  6070.24 &    18.50 &     7.89  &   -0.645  &  2.5838e+00  &  1.7596e-07  &  3.3212e-08 &  2.0388e+10 &  2.7244e+12\\
 1440 &   -5.453 &  6133.33 &    18.49 &     7.90  &   -0.656  &  2.1287e+00  &  1.4193e-07  &  2.9718e-08 &  1.7764e+10 &  2.1973e+12\\
 1476 &   -5.512 &  6174.96 &    18.49 &     7.90  &   -0.661  &  1.7659e+00  &  1.1575e-07  &  2.6162e-08 &  1.4892e+10 &  1.7919e+12\\
 1512 &   -5.578 &  6200.19 &    18.49 &     7.90  &   -0.661  &  1.5195e+00  &  9.8364e-08  &  8.0906e-08 &  1.0987e+10 &  1.5227e+12
\end{longtable}
\tablefoot{Model tabulated with $\Delta z=36$~km. The full model with $\Delta z=12$~km can be obtained electronically at the CDS (dmagele.dat).}
\end{appendix}

\end{document}